\documentclass[reprint,amsmath,amssymb,aps,prapplied,superscriptaddress]{revtex4-2}

\usepackage{amsmath,amssymb,bm}
\usepackage{graphicx}
\usepackage[hidelinks]{hyperref}
\usepackage{physics}

\begin{document}
\title{Quantum amplification with a linear quantum coupler}
\author{Changqing Wang}\email{wangchangqing@wustl.edu}
\author{Yao Lu}\email{yaolu@fnal.gov}

\affiliation{Superconducting Quantum Materials and Systems Center, Fermi National Accelerator Laboratory (FNAL), Batavia, IL 60510, USA}

\date{November 17, 2026}

\begin{abstract}
Josephson parametric amplifiers (JPAs) are essential components for quantum-limited microwave measurement, but their dynamic range is often constrained by Kerr and other pump-induced nonlinear effects. In this work, we systematically study the performance of a parametric amplifier based on the linear inductive coupler (LINC). Starting from the driven-circuit Hamiltonian, we derive the effective three-wave-mixing amplifier model, including leading flux-driven nonlinear corrections relevant to realistic operation. We analyze the resulting gain response, saturation behavior, and intermodulation distortion, comparing them with representative state-of-the-art JPA designs. Our results show that the symmetry-protected structure of the LINC suppresses static Kerr nonlinearity. Moreover, suitable circuit-parameter optimization can further reduce the dressed-squeezing and driven-Kerr effects, leading to considerable improvements in power-handling metrics. These results identify the LINC architecture as a promising platform for high-linearity Josephson parametric amplification.
\end{abstract}

\maketitle

\section{Introduction}
Quantum-limited Josephson parametric amplifiers (JPAs) enable phase-preserving amplification of microwave signals with added noise approaching half a photon in the high-gain limit, known as the standard quantum limit~\cite{caves1982quantum,roy2016introduction}. They have advanced to essential components in state-of-the-art superconducting qubit readout schemes, quantum sensing and metrology, and spin resonance detection. So far, the JPAs have been developed based on diverse resonant Josephson junction (JJ)-based devices, including the current/flux-driven superconducting nonlinear asymmetric inductive element (SNAIL)~\cite{frattini2017dipole,sivak2019kerr,sivak2020josephson,ezenkova2022broadband,joshi2025lumped,dai2025optimizing,shaw2026practical}, superconducting quantum interference device (SQUID)~\cite{yamamoto2008flux,castellanos2008amplification,krantz2013investigation,zhou2014high,eichler2014quantum,mutus2014strong,roy2015broadband,boutin2017effect,planat2019understanding,white2023readout}, and others~\cite{lecocq2021efficient,hao2026wireless}. On a separate front, significant advances have been made in traveling-wave parametric amplifiers (TWPAs) based on Josephson junctions and high-kinetic-inductance superconducting thin films~\cite{peng2022floquet,malnou2021three,macklin2015near}. However, maintaining near-quantum-limited noise over a broad bandwidth remains an important challenge.

The performance of JPAs is quantified by key figures of merit, including gain, bandwidth, added noise, dynamic range, and intermodulation distortion (IMD). Notably, the dynamic range, which represents the tolerance to signal power, poses a narrow constraint to the signal power handling capability of the state-of-the-art amplifiers. It is known that Kerr nonlinearity is the primary factor limiting the dynamic range of SNAIL- and SQUID-based JPAs, beyond which the amplification behavior becomes severely distorted~\cite{frattini2018optimizing,sivak2019kerr}. Such photon-number-dependent nonlinear behavior not only leads to gain compression at high signal power but also generates spurious tones in the presence of a multi-tone input, known as IMD, thereby limiting the multiplexing capability for multi-qubit readout applications~\cite{liu2017jpc,frattini2018optimizing,sivak2019kerr,remm2023intermodulation,kaufman2023josephson}.

In recent years, considerable efforts have focused on Josephson junction device engineering and layout optimization to mitigate unwanted nonlinearity in JPAs~\cite{frattini2018optimizing,liu2020optimizing}. It has been demonstrated that JPAs based on an array of wave mixers, such as SNAILs and SQUIDs, exhibited enhanced three-wave mixing (3WM) amplification processes and reduced Kerr nonlinearity~\cite{zhou2014high,planat2019understanding,kaufman2023josephson,kaufman2025simple,zuo2026josephson,frattini2018optimizing,sivak2020josephson}; nevertheless, the practical sizes of JJs and the presence of parasitic geometric inductance pose constraints on the number of inductor elements in the actual array layout~\cite{frattini2018optimizing}. Therefore, mitigating the nonlinear effects in individual amplifier units remains critical. Notably, current-driven SNAIL parametric amplifiers (SPAs) tailored to the static-Kerr-free point through DC flux bias~\cite{chapman2023high,eriksson2024universal} have exhibited effective suppression of static Kerr~\cite{frattini2018optimizing,sivak2019kerr,miano2022frequency}. Yet, in single-flux-biased SNAILs, this approach generally requires careful flux-bias tuning, which simultaneously modifies the resonance frequency and nonlinear coefficients. Moreover, the pump-photon-dependent driven Kerr effects emerge as the dominant source of gain compression and IMD in the desirable high-gain operation regime, thereby limiting the dynamic range of static-Kerr-free SPAs~\cite{sivak2019kerr}.

Recently, Maiti et al.~\cite{maiti2025linear} proposed a circuit for
static-Kerr-free operation that adds a linear inductive shunt to the
symmetry-protected, differentially driven SQUID of Ref.~\cite{lu2023parity}. This device, known as the linear inductive coupler (LINC), functions as a clean parametric coupler for bosonic control~\cite{maiti2025linear, maiti2026dynamically}. A circuit with a similar geometry, known as the symmetrically threaded
SQUID (STS), was previously studied as a platform for Kerr-cat
qubits~\cite{bhandari2025symmetrically}. More recently, Yanik et al.
studied a variant with a linear inductive shunt and predicted
near-quantum-limited amplification in the weak-signal regime at a
static-Kerr-free bias~\cite{yanik2026flux}. However, the power-handling capability and IMD property of LINC-based JPAs remain largely unexplored, as the effects of dressed squeezing and flux-driven Kerr nonlinearities on finite-signal performance have not been clarified.

In this study, we systematically investigate the dynamic range of a LINC-based parametric amplifier (LPA), through a comprehensive characterization of its gain compression
and IMD properties. Our analysis employs a realistic model that incorporates the leading flux-pump-induced nonlinear corrections, including photon-number-dependent dressed squeezing and
driven Kerr nonlinearity. By implementing circuit
design strategies, including optimization of the linear inductance with respect to the total capacitance and the use of series LINC
arrays, we further suppress the residual nonlinearities while preserving
the desired small-signal gain. Our results show improved power handling in the matched SQUID comparison and higher IIP$_3$ than the SNAIL benchmarks considered here.

\section{Ideal model of a LINC parametric amplifier}

We consider the driven operation of a LINC coupler, characterized by the charging energy $E_C$, the Josephson energy $E_J$ of each junction, and the inductive energy $E_L$ of the shunting inductor. The Hamiltonian takes the form~\cite{maiti2025linear}
\begin{align}
    H_{\text{LINC}} & = 4E_c\hat{n_c}^2+\frac{E_L}{2}\hat{\theta_c}^2-2E_J\cos\phi_d\cos\hat{\theta}_c,\label{1}
\end{align}
where $\hat{n_c}$ and $\hat{\theta_c}$ are the conjugate charge and common mode phase operators, respectively. We introduce the instantaneous ladder operators $\hat{c}$ and $\hat{c}^{\dagger}$ through $\hat{\theta}_c=\theta_{\mathrm{zpf}}\left(\hat{c}+\hat{c}^{\dagger}\right)$, with the flux-dependent zero-point-fluctuation phase defined as \begin{align}
    \theta_{\mathrm{zpf}}=\left(\frac{2E_C}{E_L+2E_J\cos\phi_d}\right)^{1/4}.\label{theta_zpf}
\end{align}

Analogous to the driven operation of LINC, the LPA operates under an external flux $\phi_d$ with a DC bias $\phi_{\text{DC}}$ and an additional AC modulation at frequency $\omega_d$, i.e., $\phi_d=\phi_{\text{DC}}+\phi_{\text{ac}}(t)=\phi_{\text{DC}}+\abs{\phi_{\text{ac}}}\cos(\omega_d t + \phi_0)$. By setting the operation bias at $\pi/2$, the static nonlinear term in the Hamiltonian vanishes. Expanding Eq.~(\ref{1}) to the second order in the phase operator then yields a linear flux-driven Hamiltonian
\begin{align}
    H_{\text{LINC}}[\phi_{DC}=\frac{\pi}{2}] & =  \left(\omega_{L}+2g_{22}\abs{\phi_{\mathrm{ac}}}^2\right) \hat{c}^{\dagger} \hat{c} \nonumber\\&+ \left(2g_{21}\phi_{ac}(t)+2g_{22}\phi_{\mathrm{ac}}^2(t)\right) \left(\hat{c}^2 + \hat{c}^{\dagger2} \right),
\end{align}
with the resonance frequency $\omega_L=\sqrt{8E_C E_L}$, and the parametric strengths $g_{21,22}$ arising from the modulation of the potential energy (\hyperref[AppendixA]{Appendix A}).

To enable parametric amplification, we apply an AC flux at a frequency $\omega_d$ near $2\omega_L$. After applying the rotating-wave approximation (RWA) and dropping the counter-rotating wave terms, we obtain the Hamiltonian in the rotating frame
\begin{align}
    H_{\text{LINC}}[\phi_{DC}=\frac{\pi}{2}] &= \left(\Delta+2g_{22}\abs{\phi_{ac}}^2\right) \hat{c}^{\dagger} \hat{c} \nonumber\\&+ g_{21}\abs{\phi_{ac}} \left(e^{i\phi_0}\hat{c}^2 + e^{-i\phi_0}\hat{c}^{\dagger2} \right),
\end{align}
with $\Delta=\omega_L-\omega_d/2$. Finally, the Hamiltonian can be written in a generic single-mode squeezing form
\begin{equation}
    H_{\text{LINC}} = \hbar\delta \hat{c}^{\dagger} \hat{c}
    +\frac{i \hbar}{2}\left( \epsilon \hat{c}^{\dagger 2} -\epsilon^{*} \hat{c}^2\right),\label{H_linear}
\end{equation}
with $\delta=\Delta+2g_{22}\abs{\phi_{ac}}^2$ and $\epsilon=-i2g_{21}\abs{\phi_{ac}}e^{-i\phi_0}$. Hence, the expansion up to quadratic-phase terms yields the model of a perfectly linear parametric amplifier.

In a parametric amplifier working scheme, the signal is injected through an external channel that couples to the LINC ($a_{\rm in}$) by the coupling rate $\gamma_c$. Additionally, the intrinsic loss channel ($b_{\rm in}$) with decay rate $\gamma_0$ introduces vacuum noise.
The evolution of $\hat{c}$ and $\hat{c}^{\dagger}$ thereby follows the quantum Langevin equations (QLE)
\begin{align}
    \frac{d\hat{c}}{dt}&=(-i\delta-\frac{\gamma}{2}) \hat{c} + \epsilon \hat{c}^{\dagger} + \sqrt{\gamma_c} a_{\rm in}+\sqrt{\gamma_0} b_{\rm in},\label{EM1}\\
    \frac{d\hat{c}^{\dagger}}{dt}&=(i\delta-\frac{\gamma}{2}) \hat{c}^{\dagger} + \epsilon^* \hat{c} + \sqrt{\gamma_c} a_{\rm in}^{\dagger}+\sqrt{\gamma_0} b_{\rm in}^{\dagger},\label{EM2}
\end{align}
with the total dissipation rate
$\gamma=\gamma_0+\gamma_c$.
Solving Eqs.~(\ref{EM1}) and (\ref{EM2}) in the frequency domain and combining the input-output relation yields the gain spectrum for an ideal LPA (\hyperref[AppendixB]{Appendix B})
\begin{align}
    G_s(\omega) = \frac{\begin{gathered}
    \left[\omega^2-\delta^2+\abs{\epsilon}^2+\gamma(\gamma_c-\gamma_0)/4\right]^2\\
    +\left(\gamma_0\omega-\gamma_c\delta\right)^2
    \end{gathered}}{\left[\omega^2-\delta^2+\abs{\epsilon}^2-\gamma^2/4\right]^2+\gamma^2\omega^2}.
\end{align}

\section{Effects of higher-order nonlinearities}
The ideal model introduced above is unable to characterize the finite-signal response of an LPA. To understand its power-handling capabilities, we now derive a comprehensive model of LPAs that incorporates higher-order nonlinear effects induced by flux-driven effects. Retaining terms up to fourth order in the phase operator, the Hamiltonian becomes
\begin{align}
    H_{\text{LINC}}[\phi_{DC}=\frac{\pi}{2}]&=\left(\omega_{L}+2g_{22}\abs{\phi_{ac}}^2\right) \hat{c}^{\dagger} \hat{c}\nonumber\\&+ \left(2g_{21}\phi_{ac}(t)+2g_{22}\phi_{ac}^2(t)\right) \left(\hat{c}^2 + \hat{c}^{\dagger2} \right)\nonumber\\&+(g_{41}\phi_{ac}(t)+g_{42}\phi_{ac}^2(t))\left(\hat{c}+\hat{c}^{\dagger}\right)^4.
\end{align}
Under the RWA, the Hamiltonian in the frame rotating at $\omega_d/2$ reduces to the form (see \hyperref[AppendixA]{Appendix A})
\begin{align}
    H_{\text{LINC}}[\phi_{DC}=\frac{\pi}{2}] &= \hbar\delta_{\text{eff}} \hat{c}^{\dagger} \hat{c}+\frac{i \hbar}{2}\left( \epsilon_1 \hat{c}^{\dagger 2} -\epsilon_1^{*} \hat{c}^2\right)\nonumber\\&+\frac{i \hbar}{2}\left( \epsilon_2 \hat{c}^{\dagger 2} \hat{c}^{\dagger}\hat{c} -\epsilon_2^{*} \hat{c}^{\dagger}\hat{c} \hat{c}^2\right)\nonumber\\&+\frac{1}{2}\hbar K_{eff}\hat{c}^{\dagger 2} \hat{c}^{2}\nonumber\\&+\frac{1}{24}\hbar K_{\text{eff}}(e^{2i\phi_0}c^4+e^{-2i\phi_0}c^{\dagger4}),\label{10}
\end{align}
which includes several terms: the effective detuning containing the AC Stark shift
\begin{equation}    \delta_{\text{eff}}=\Delta+2g_{22}\abs{\phi_{ac}}^2+6g_{42}|\phi_{ac}|^2,
\end{equation}
the intrinsic squeezing with strength
\begin{equation}    \epsilon_1=-i(2g_{21}+6g_{41})|\phi_{ac}|e^{-i\phi_0},
\end{equation}
the dressed squeezing~\cite{frattini2024doublewell,yanik2026flux} with strength
\begin{equation}
    \epsilon_2=-4ig_{41}|\phi_{ac}|e^{-i\phi_0},
\end{equation} and the AC-flux-driven Kerr with coefficient
\begin{equation}
    K_{\text{eff}}=6g_{42}|\phi_{ac}|^2.
\end{equation}

\begin{figure}[!t] 	\centering\includegraphics[width=\columnwidth]{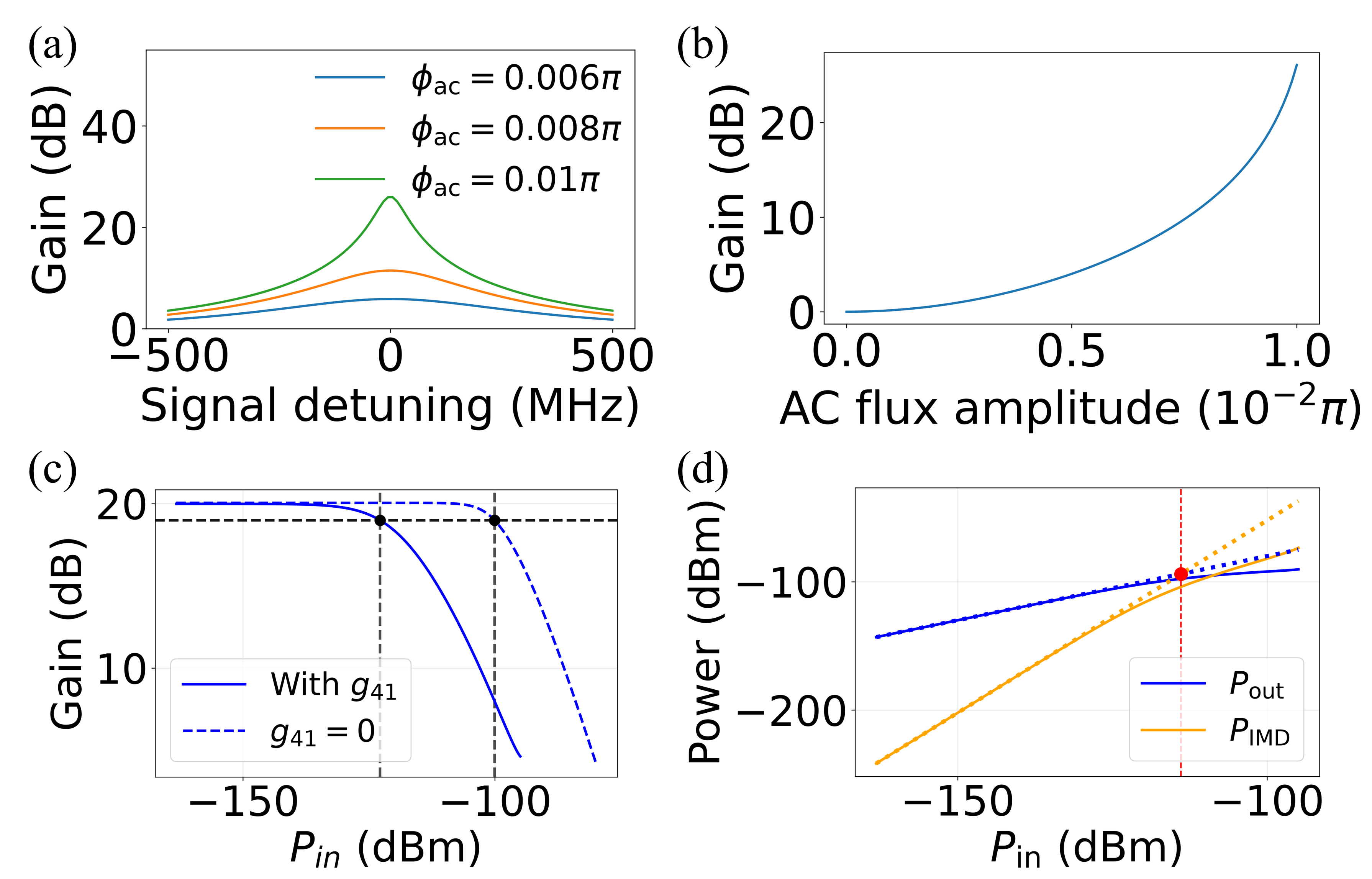}
	\caption{\normalsize Properties of LPA. (a) Small-signal gain as a function of the signal detuning for different AC flux amplitudes $\phi_{\text{ac}}$. (b) Small-signal gain as a function of $\phi_{\text{ac}}$. (c) Gain as a function of input signal power with the dressed squeezing included (solid) and excluded (dashed). The black dots mark the $P_{\text{1dB}}$ points. Parameters: $E_L/h=2.64$\,THz, $E_C/h=2.32$\,MHz, $E_J/h=2.29$\,THz, $\gamma/2\pi=200$\,MHz, $|\phi_{\text{ac}}|/\pi=9.5\times10^{-3}$, $\Delta/2\pi=1.17$\,MHz. (d) The output power at the signal frequency (solid blue) and the intermodulation power (solid orange) as a function of the input power. IIP$_{3}$ is determined by the intersection (red dot) of the lower-power asymptotes (dashed curves).}  \label{Fig1}
\end{figure}

To analyze the field dynamics under the effect of the quartic nonlinear terms, we rewrite the QLEs following the Hamiltonian given by Eq.~(\ref{10})
\begin{align}
    \frac{d\hat{c}}{dt}&=(-i\delta_{\text{eff}}-\frac{\gamma}{2}) \hat{c} + \epsilon_1 \hat{c}^{\dagger} + \frac{3}{2}\epsilon_2 \hat{c}^{\dagger2}\hat{c} - \frac{1}{2}\epsilon_2^* \hat{c}^{3} \nonumber\\&-i K_{\text{eff}} \hat{c}^{\dagger}\hat{c}^2-i\frac{1}{6}K_{\text{eff}}e^{-2i\phi_0}c^{\dagger3} +\sqrt{\gamma_c} \hat{a}_{\rm in}+\sqrt{\gamma_0} \hat{b}_{\rm in},\label{EM3}\\
    \frac{d\hat{c}^{\dagger}}{dt}&=(i\delta_{\text{eff}}-\frac{\gamma}{2}) \hat{c}^{\dagger} + \epsilon_1^* \hat{c} + \frac{3}{2}\epsilon_2^* \hat{c}^{\dagger}\hat{c}^2 - \frac{1}{2}\epsilon_2 \hat{c}^{\dagger3} \nonumber\\&+i K_{\text{eff}} \hat{c}^{\dagger2}\hat{c}+i\frac{1}{6}K_{\text{eff}}e^{2i\phi_0}c^{3} + \sqrt{\gamma_c} \hat{a}_{\rm in}^{\dagger}+\sqrt{\gamma_0} \hat{b}_{\rm in}^{\dagger}.\label{EM4}
\end{align}
Further, JPAs typically operate in the strongly overcoupling regime, leading to the approximation $\gamma\approx\gamma_c$. Following the semiclassical harmonic balance approach applied for SPAs~\cite{frattini2018optimizing,sivak2019kerr}, we can solve the QLEs and obtain the photon-number-dependent gain
\begin{align}
    G & =\abs{\frac{i\gamma}{D}\left(-\omega-\Delta_s-\frac{i\gamma}{2}\right) -1}^2,\label{G}
\end{align}
where
\begin{align}
    D&=-\omega^2+\Delta_s \Delta_i + \omega(\Delta_i-\Delta_s)+\gamma^2/4-4g_{\text{eff,i}}^*g_{\text{eff,s}}\nonumber\\&-i\gamma(\omega+\frac{\Delta_s-\Delta_i}{2}).
\end{align}
$\Delta_{s,i}$ and $g_{\text{eff},s,i}$ are photon-number-dependent detuning and squeezing amplitudes, respectively (see \hyperref[AppendixB]{Appendix B}).

In the weak-signal regime, the LPA operates as a linear JPA with a small-signal gain
\begin{align}
    G_0 =1+\frac{4\gamma^2\abs{g_{\text{eff}}}^2}{\left[-\omega^2+\delta_{\text{eff}}^2+\frac{\gamma^2}{4}-4\abs{g_{\text{eff}}}^2\right]^2+\gamma^2\omega^2}.\label{G0}
\end{align}
Fig.~\ref{Fig1}(a) shows the gain spectrum of a flux-driven LPA under realistic parameters. The small-signal gain increases with $\phi_{\text{ac}}$, allowing a gain of 20\,dB to be achieved with a moderate flux-drive amplitude [Fig.~\ref{Fig1}(b)].

At higher signal power, however, the dressed squeezing and driven Kerr become significant, giving rise to gain compression and third-order IMD. The gain at each input signal power can be found by numerically solving the EOMs using Eqs.~(\ref{eq:geff})-(\ref{eq:Deltas}) in \hyperref[AppendixB]{Appendix B}. The resulting 1\,dB gain compression point, $P_{1\text{dB}}$, defined as the input signal power at which the gain drops by 1\,dB from its small-signal value, is found to be $-123$\,dBm.

While previous work identified Kerr nonlinearity as the main cause of amplifier deviations in the weak-signal regime~\cite{yanik2026flux}, our results find that the dressed squeezing characterized by the $g_{41}$ is the primary source of gain compression [Fig.~\ref{Fig1}(c)]. By comparison, gain compression and IMD in the charge-driven SPAs studied in Refs.~\cite{frattini2018optimizing,sivak2019kerr} are primarily attributed to static or driven Kerr nonlinearity.

Furthermore, it is well known that the Kerr nonlinearity can be reliably estimated from the intermodulation behavior of the SPAs~\cite{frattini2018optimizing,sivak2019kerr}. Here we find that the IMD property of an LPA is determined by both the driven Kerr and dressed squeezing (see \hyperref[AppendixC]{Appendix C}).
With two near-resonance signals at frequencies
$\omega_{s1,s2}$ and equal input power $P_{\mathrm{in}}$
per tone, the output signal power and the IMD power at
$2\omega_{s1}-\omega_{s2}$ and $2\omega_{s2}-\omega_{s1}$
scale as $P_{\mathrm{in}}$ and $P_{\mathrm{in}}^3$,
respectively, in the weak-signal regime at fixed pump.
The intersection of their low-power asymptotes defines
the input-referred third-order intercept point
($\text{IIP}_3$) [Fig.~\ref{Fig1}(d)].
For $\delta_{\mathrm{eff}}=0$ and distinct signal, idler,
and IMD frequencies with detunings small compared with
the gain bandwidth, we obtain
(see \hyperref[AppendixC]{Appendix C} for the derivation)
\begin{align}
    \text{IIP}_3 = \left|\frac{\frac{\gamma^2}{4}-|\varepsilon_1|^2}{\frac{\gamma}{2}C_A+\epsilon_1C_B}\right|\frac{\sqrt{G_0}}{(\sqrt{G_0}+1)^3}\hbar\omega_L \gamma,
\end{align}
where
\begin{align}
C_A&=
-iK_{\text{eff}}(1+2|r|^2)-\frac{i}{2}K_{\text{eff}}e^{-2i\phi_0}r^2
\nonumber\\
&+\frac{3}{2}\varepsilon_2 r(2+|r|^2)-\frac{3}{2}\varepsilon_2^*r^*,\\
C_B&=
iK_{\text{eff}}r(2+|r|^2)+\frac{i}{2}K_{\text{eff}}e^{2i\phi_0}r^*
\nonumber\\
&+\frac{3}{2}\varepsilon_2^*(1+2|r|^2)-\frac{3}{2}\varepsilon_2 r^2,\\
r&=
\frac{\varepsilon_1^*}
{\gamma/2}.
\end{align}

While dressed squeezing is an additional concern for LPAs, the following section sheds light on how to mitigate this effect through device optimization.

\section{Optimization of the LINC parametric amplifier}
To quantify the impact of dressed squeezing in the amplification process, we evaluate the ratio between the dressed squeezing and the ordinary squeezing amplitudes
\begin{equation}
    R_1 = \frac{\abs{\epsilon_2}}{\abs{\epsilon_1}} \approx \frac{1}{3}\sqrt{\frac{2E_C}{E_L}}=\frac{2e^2}{3\hbar}\sqrt{\frac{L}{C}}.
\end{equation}
The above equation indicates that maximizing $E_L/E_C$ can effectively reduce the dressed squeezing effect and mitigate gain compression.

On the other hand, the impact of the driven Kerr nonlinearity can be quantified by the ratio between the driven Kerr coefficient and the squeezing amplitude
\begin{align}
    R_2&=\frac{K_{\text{eff}}}{|i\epsilon_1|/2} \approx \frac{6g_{42}|\phi_{ac}|^2}{|g_{21}||\phi_{ac}|}=\frac{4|g_{21}||\phi_{ac}|}{E_L}.
\end{align}
The approximation is valid for $E_C \ll 2E_L$. Thus, increasing $E_L$ also suppresses the driven Kerr nonlinearity.

When tuning the circuit parameters, the charging energy $E_C=\omega_L^2/8E_L$ must be adjusted to maintain a fixed resonant frequency. In addition to resonance matching, we choose a 20\,dB gain at the small-signal limit ($G_0$) for fair comparison with conventional JPAs. Thus, we maintain the condition
\begin{equation}
    2g_{\text{eff}}=-\frac{E_J}{2E_L}\omega_L|\phi_{ac}|e^{-i\phi_0} =\text{const}.
\end{equation}
Hence, in our parameter tuning, the reduction in $g_{\text{eff}}$ due to larger $E_L$ is compensated by increasing $E_J$ in proportion.

\begin{figure}[!htb] 	\centering\includegraphics[width=\columnwidth]{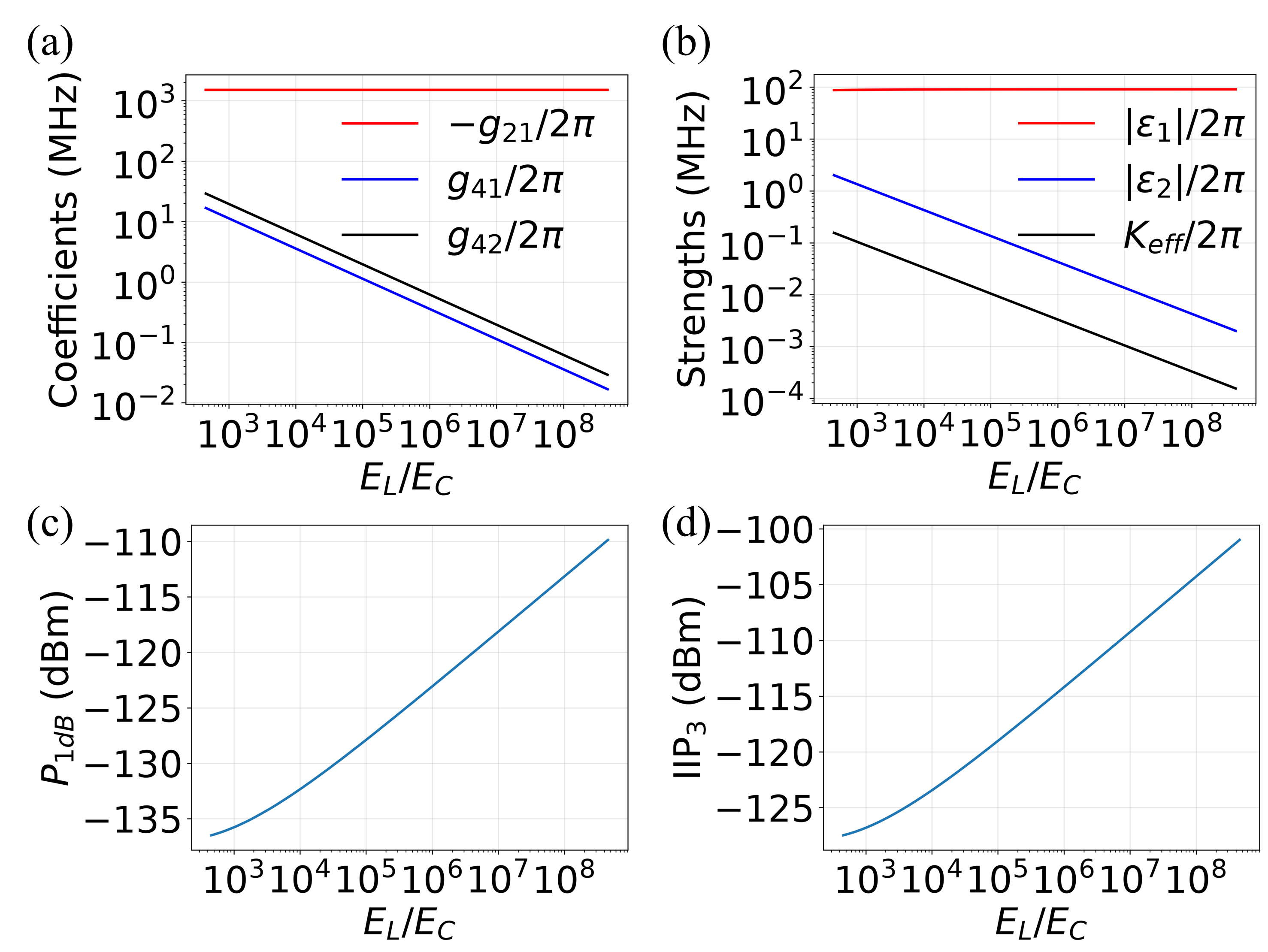}
	\caption{\normalsize Optimization of LPA figures of merit by tuning $E_L/E_C$. (a) Nonlinear coefficients, (b) wave mixing strengths, (c) $P_{1\text{dB}}$, and (d) $\text{IIP}_3$ as a function of $E_L/E_C$, for an LPA with 20\,dB small-signal gain. $E_J$ is increased proportionally to $E_L$.
    }\label{Fig2}
\end{figure}

As shown by Fig.~\ref{Fig2}(a)(b), both the dressed squeezing and Kerr nonlinearity are dramatically suppressed by increasing $E_L/E_C$, with $G_0=20\,$dB.
As a result, the gain compression of LPAs is significantly mitigated by increasing $E_L/E_C$, as evidenced by the improvement of $P_{\text{1dB}}$ [Fig.~\ref{Fig2}(c)].
Moreover, a larger $E_L/E_C$ simultaneously improves
$\text{IIP}_3$ [Fig.~\ref{Fig2}(d)]. A comparison with
reported SNAIL-array performance is presented in
Table~\ref{tab:table1}.

\section{LINC-Array-based parametric amplifier}
\begin{figure}[!t] 	\centering\includegraphics[width=\columnwidth]{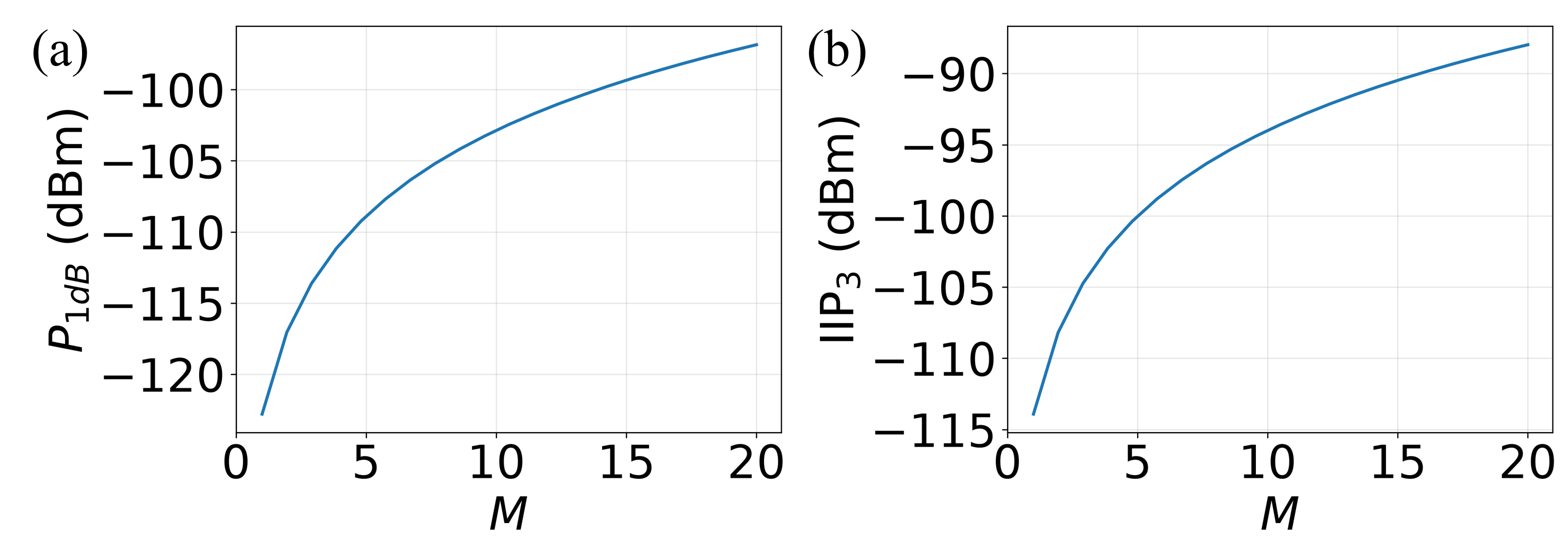}
	\caption{\normalsize LINC-array-based LPAs. (a) $P_{1\text{dB}}$ and (b) $\text{IIP}_3$ as a function of the number of LINCs ($M$), with 20\,dB small-signal gain.}\label{Fig3}
\end{figure}

Motivated by the array architecture employed in SNAIL- and SQUID-based parametric amplifiers, we propose a LINC-array-based parametric amplifier comprising multiple LINCs connected in series~\cite{eichler2014controlling,frattini2018optimizing,maiti2025linear} to reduce the dressed squeezing and flux-driven Kerr nonlinearity. To maintain a fixed operating frequency for an $M$-LINC array, we fix the total capacitance and reduce the linear and Josephson inductance in each LINC unit to $1/M$ of their original values, such that $\tilde{E_L}=ME_L$ and $\tilde{E_J}=ME_J$.
For identical LINCs subject to the same flux drive,
the total energy of the $M$-LINC array yields the Hamiltonian
\begin{align}
    H_{\text{LINC array}} & = 4E_c\hat{n_c}^2+\frac{M^2E_L}{2}\left(\frac{\hat{\theta_c}}{M}\right)^2\nonumber\\&-2M^2E_J\cos\phi_d\cos(\frac{\hat{\theta}_c}{M}).\label{27}
\end{align}
Since the quadratic term remains invariant, $\theta_{\text{zpf}}$ takes the same form as Eq.~(\ref{theta_zpf}); so do $g_{2n}$ coefficients. In contrast, the higher-order nonlinear terms exhibit $M$-dependent behaviors.
By expanding Eq.~(\ref{27}) with respect to $\hat{\theta_c}$
\begin{align}
    H_{\text{LINC array}} & = 4E_c\hat{n_c}^2+\frac{E_L}{2}\hat{\theta_c}^2\nonumber\\&+E_J\cos\phi_d\hat{\theta}_c^2-\frac{E_J}{12}\cos\phi_d\frac{\hat{\theta}_c^4}{M^2}+\cdots,
\end{align}
we find that the quartic phase terms scale as $1/M^2$, leading to a $M^2$ reduction of $g_{41}$ and $g_{42}$, thereby suppressing both the dressed squeezing and flux-driven Kerr nonlinearity. As a result, increasing $M$ improves both $P_{\text{1dB}}$ and $\text{IIP}_3$ (Fig.~\ref{Fig3}), which reach the levels of $-97$\,dBm and $-88$\,dBm, respectively, hence substantially enhancing the dynamic range.

\section{Comparison with SQUID and SNAIL-based parametric amplifiers}
To further benchmark the advantages of the LPA in nonlinearity suppression, we numerically compare its \(P_{1\mathrm{dB}}\) and IIP3 with those of a conventional SQUID-based JPA. The Hamiltonian of the symmetric SQUID under RWA in the rotating frame takes the form
\begin{equation}
\begin{aligned}
H_{\mathrm{SQUID}}
={}&\hbar\delta_{\mathrm{eff}}^S\hat{c}^{\dagger}\hat{c}
+\frac{i\hbar}{2}
\left(
\epsilon_1^S\hat{c}^{\dagger 2}
-\epsilon_1^{S*}\hat{c}^2
\right)+\frac{\hbar}{2}K_{\mathrm{eff}}^S
\hat{c}^{\dagger 2}\hat{c}^2,
\end{aligned}
\end{equation}
where
\begin{align}
\delta_{\mathrm{eff}}^S
&=\Delta^S+2g_{22}^S|\phi_{\mathrm{ac}}^S|^2-E_C^S,\\
\epsilon_1^S
&=-2ig_{21}^S|\phi_{\mathrm{ac}}^S|e^{-i\phi_0^S},\\
K_{\mathrm{eff}}^S
&=-E_C^S.
\end{align}
Unlike the LPA operating at its static-Kerr-free bias point, the SQUID exhibits an intrinsic quartic nonlinearity, giving rise to a static Kerr coefficient \(K^{S}_{\text{eff}}=-E_C\). For the SQUID benchmark, we retain static Kerr as the only quartic
term in the effective Hamiltonian. For a fair comparison, we match the resonance frequencies, charging energies, and parametric amplification coefficients of the two amplifiers, while independently adjusting their pump amplitudes to achieve the same small-signal gain (see \hyperref[AppendixD]{Appendix D} for details). The gain of the SQUID-based JPA is derived using the harmonic balance approach (\hyperref[AppendixD]{Appendix D}).

\begin{figure}[!t] 	\centering\includegraphics[width=\columnwidth]{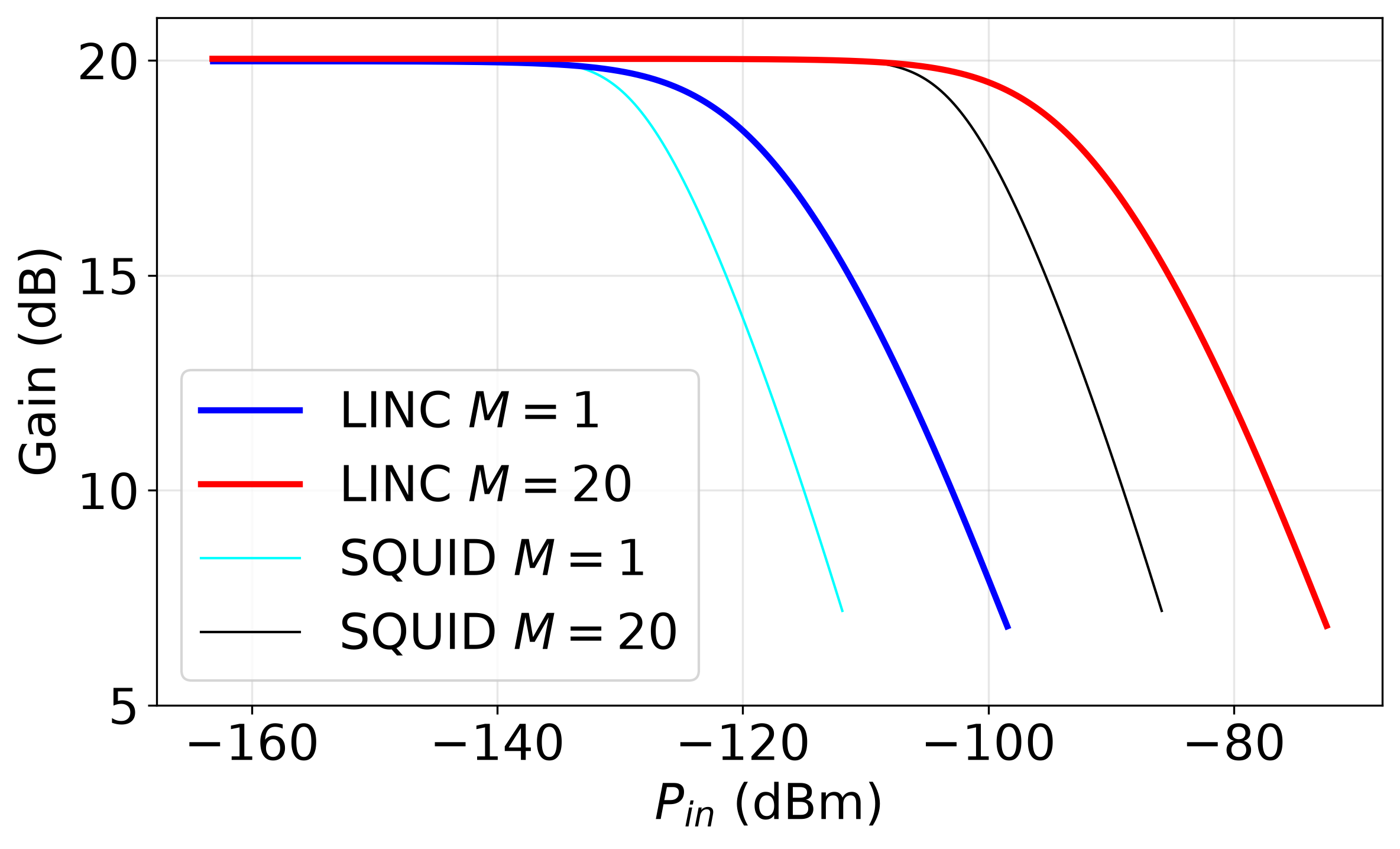}
	\caption{\normalsize Comparison of LINC- and SQUID-based parametric amplifiers gain as a function of the input power $P_{\text{in}}$, for an LPA, a LINC-array-based LPA, a SQUID-based JPA, and a SQUID-array-based JPA. For the SQUID-based JPA, we choose $E_C^S = E_C$, $E_J^S=E_L$, $\phi_{DC}^S=\pi/3$.
    }\label{Fig4}
\end{figure}

We numerically evaluate the gain as a function of the input power for an LPA, a LINC-array-based LPA, a SQUID-based JPA, and a SQUID-array-based JPA (Fig.~\ref{Fig4}). The result shows that the LPA exhibits superior gain compression and IMD properties, with $P_{1\text{dB}}=-123$\,dBm and $\text{IIP}_3=-114$\,dBm; the SQUID-based JPA exhibits a relatively smaller dynamic range with $P_{1\text{dB}}=-129$\,dBm and $\text{IIP}_3=-120$\,dBm. As noted in the previous section, further improvement is possible by implementing an array architecture. Notably, the parametric amplifier employing a 20-LINC-array achieves $P_{1\text{dB}}=-97$\,dBm and $\text{IIP}_3=-88$\,dBm, demonstrating superior power-handling capability and lower IMD compared with its 20-SQUID-array counterpart, which exhibits $P_{1\text{dB}}=-103$\,dBm and $\text{IIP}_3=-94$\,dBm.

In addition to conventional symmetric Josephson-junction architectures such as SQUIDs, asymmetric circuits enabling three-wave mixing, such as SNAILs, have been developed to suppress Kerr nonlinearity in JPAs. Notably, the charge-driven SPAs have demonstrated outstanding JPA performance at the operating point that enables static-Kerr cancellation. Nevertheless, the charge-driven Kerr nonlinearity persists, limiting the dynamic range. To compare the LPA with SPA, we benchmark the LPA predictions at 20\,dB small-signal gain and 7\,GHz against reported SPA performance. Our analysis adopts an effective model of SPA that incorporates the second-order perturbation arising from three-wave mixing (see \hyperref[AppendixE]{Appendix E} for details). The near-resonance signal behavior of the SPA can be described by an effective Hamiltonian
\begin{align}
    H_{\text{SNAIL,eff}}/\hbar &= - \Delta_b \hat{b}^\dagger \hat{b} +  ( g \hat{b}^{\dagger2} + g^* \hat{b}^2)+\frac{K}{2}\hat{b}^{\dagger 2}\hat{b}^2,
\end{align}
where
\begin{align}
    \Delta_b (n_p) &= \Delta-(\frac{32}{3}g_4-28\frac{g_3^2}{\omega_a})n_{p},\\
    g(\alpha_p)&=2g_3\alpha_p,\\
    K(n_p)&=12(g_4-5\frac{g_3^2}{\omega_a})-\frac{256}{3}\frac{g_4^2 n_p}{\omega_a}.\label{Kerr}
\end{align}

In conventional SPAs, a substantial static Kerr nonlinearity arises from the $12g_4$ term. However, applying an external flux bias allows tuning of $g_3$ to balance the $-60\frac{g_3^2}{\omega_a}$ term in Eq.~(\ref{Kerr}) against the $12g_4$ term, thereby achieving effective static Kerr cancellation \cite{sivak2019kerr}. The signal-power saturation of SPA near the Kerr-cancellation point is dominated by an unbalanced pump-photon-number-dependent term $\frac{256}{3}\frac{g_4^2n_p}{\omega_a}$ in Eq.~(\ref{Kerr}).

Like SQUIDs and LINCs, arraying SNAILs substantially reduces the undesirable nonlinearity. For an SPA composed of an array of $M$ SNAILs, the nonlinear coefficient scales as $g_3\rightarrow g_3/\sqrt{M}$, $g_4\rightarrow g_4/M$. As a result, reducing the ratio between $g_4$ and $g_3$ suppresses the driven Kerr nonlinearity while preserving gain~\cite {frattini2018optimizing,sivak2019kerr}.

The primary constraint on the performance of SNAIL-based SPAs in the high-gain regime is the pump-photon-dependent driven Kerr nonlinearity, among other higher-order nonlinear effects, even when one implements the array architecture and operates the device around the static-Kerr-free point.

To benchmark LPAs against state-of-the-art SPAs, we summarize their figures of merit in Table~\ref {tab:table1}. It is worth noting that the LINC array yields around 6-14\,dB improvement of $\text{IIP}_3$ compared with its SPA counterpart, while $P_\text{1dB}$ is comparable to the SNAIL-array benchmarks listed in Table~\ref{tab:table1}.
This improvement combines the symmetry-protected cancellation
of static Josephson nonlinearity with the suppression of
residual pump-induced nonlinearities through circuit design
and arraying. Hence, the LPA is a parametric amplifier with superior power-handling and IMD properties.

\begin{table}[t]
\centering
\caption{Comparison of power-saturation metrics for SPA and LPA architectures.}
\begin{tabular}{lcc}
\hline\hline
\textbf{Amplifier Type} & \textbf{$P_{\mathrm{1dB}}$ (dBm)} & \textbf{IIP$_3$ (dBm)} \\
\hline
SNAIL-array\cite{frattini2018optimizing} & $\leq$-100 & -96  \\
SNAIL-array\cite{sivak2019kerr} & -102 & -94  \\
SNAIL-array\cite{ezenkova2022broadband} & -97 & N/A  \\
SNAIL-array\cite{joshi2025lumped} & $\leq$-100 & -102  \\
Single-LINC & -123 & -114 \\
20-LINC-array & -97 & -88 \\
\hline\hline
\end{tabular}
\label{tab:table1}
\end{table}

\section{Conclusion}

In this study, we develop a comprehensive model for flux-driven LPAs starting from the driven-circuit Hamiltonian. We investigate the effects of leading higher-order nonlinear corrections, such as dressed squeezing and the AC-flux-driven Kerr effect. By solving QLEs using the harmonic balance approach, we obtain key finite-signal amplifier properties, including gain compression and IMD, in the realistic operating regime. Our study reveals that increasing the ratio $E_L/E_C$ as well as implementing an array architecture are practical routes to suppress the residual nonlinear response and improve power-handling metrics such as $P_{1\mathrm{dB}}$ and IIP3. We further demonstrate the advantages of LPAs over conventional architectures by numerically comparing the LPA performances with SQUID-based JPAs under comparable operating conditions and benchmarking LPAs against the reported performance of SNAIL-based JPAs.

Our study suggests that the LINC architecture is a promising platform for high-ideality JPAs, with outstanding power-handling capability and IMD protection. Those merits position it as an ideal candidate for
parametric amplifiers for multi-tone or broadband input, which is essential for multiplexed readout of superconducting qubits. Notably, the advantage of LINC arises not only from its ideal linearity in the DC operation, but also
from its significant suppression of flux-driven Kerr.

As a final note, we clarify the distinction between our findings and those of a recent study that employs a symmetrically threaded SQUID (STS) for parametric amplification~\cite{yanik2026flux}. The work by Yanik et al. primarily characterizes the squeezing, small-signal gain, and quantum efficiency of an ideal Kerr-free STS amplifier under pump-driven vacuum or weak-probe conditions; the driven Kerr was neglected, and the impact of dressed-squeezing was found negligible in their regime. In contrast, our study covers the large-signal nonlinear regime, where gain compression becomes the essential limitation on readout signal amplification; the driven Kerr nonlinearity is incorporated into our model, and the large intracavity photon population makes photon-number-dependent dressed-squeezing significant.

\section*{Acknowledgments}
The authors thank Gangqiang Liu, Nicholas Frattini, Florent Lecocq,
Shyam Shankar, and Xinyuan You for helpful discussions. This work was supported by the U.S. Department of Energy, Office of Science, National Quantum Information Science Research Centers, Superconducting Quantum Materials and Systems Center (SQMS), under Contract No. 89243024CSC000002. Fermilab is operated by Fermi Forward Discovery Group, LLC under Contract No. 89243024CSC000002 with the U.S. Department of Energy, Office of Science, Office of High Energy Physics. Y.L. acknowledges support from the U.S. Department of Energy, Office of Science, Office of High Energy Physics, through the Early Career Research Program.
\appendix

\section{Derivation of the LPA Hamiltonian}\label{AppendixA}

\setcounter{figure}{0}
\def\thefigure{A\arabic{figure}}
We start with the Hamiltonian of the driven operation of a LINC biased at an arbitrary operating point $\phi_{\text{DC}}$ \cite{maiti2025linear}
\begin{align}
    H_{\text{LINC}} & = 4E_c\hat{n_c}^2+\frac{E_L}{2}\hat{\theta_c}^2-2E_J\cos\phi_d\cos\hat{\theta}_c.\label{app_1}
\end{align}
where the flux contains a DC bias and an AC flux at the frequency $\omega_d$, i.e., $\phi_d=\phi_{\text{DC}}+\phi_{\text{ac}}=\phi_{\text{DC}}+\abs{\phi_{\text{ac}}}\cos(\omega_d t + \phi_0)$.
Under the bosonic basis $\hat{\theta}_c=\theta_{\text{zpf}}\left(\hat{c}+\hat{c}^{\dagger}\right)$, the flux-driven zero-point-fluctuation phase is given by
\begin{align}
    \theta_{\text{zpf}}=\left(\frac{2E_C}{E_L+2E_J\cos\phi_d}\right)^{1/4}.\label{A2}
\end{align}

To evaluate the parametric wave mixing processes, we expand $\cos(\phi_d)$ and all $\theta_{\text{zpf}}$-dependent terms of the potential energy in Taylor series around $\phi_{\text{DC}}$
\begin{align}
    U_{\text{LINC}} &= \sum_{m\in \text{even}, n} g_{mn}[\phi_{DC}]\phi_{ac}^n(t)\left(\hat{c}+\hat{c}^{\dagger}\right)^m,
\end{align}
with
\begin{align}
    g_{mn}[\phi_{\mathrm{dc}}] &=\frac{1}{m! n!} \sum_{k=0}^{n} \binom{n}{k} u_{m,n-k} \frac{\partial^{k} \theta_{\mathrm{ZPF}}^{m}}{\partial \phi_{d}^{k}} \bigg|_{\phi_{d} = \phi_{\mathrm{DC}}},\nonumber\\
    u_{m,k}&= \frac{\partial^{m+k} U_{\mathrm{LINC}}}{\partial \hat{\theta}_c^m \partial \phi_d^k}
\Bigg|_{\hat{\theta}_c = 0}.\label{gmn}
\end{align}
For \(m=2\), the harmonic potential coefficient is \(u_{2,0}\theta_{\mathrm{zpf}}^2/2=\omega_0/4\), giving \(g_{21}=\omega_0'/4\) and \(g_{22}=\omega_0''/8\), where the derivatives are evaluated at the DC bias. Meanwhile, the kinetic and quadratic potential energies contribute equally to the harmonic level spacing. Including both kinetic and potential contributions, the leading parametric coupling under the DC-bias-oscillator basis has coefficient \(u_{2,1}\theta_{\mathrm{zpf}}^2/2=\omega_0'/2=2g_{21}\), and the cycle-averaged second-order correction to instantaneous frequency is \(2g_{22}|\phi_{\text{ac}}|^2\). The Hamiltonian expanded to the second power of $\abs{\phi_{\text{ac}}}$ and $\left(\hat{c}+\hat{c}^{\dagger}\right)$ (i.e., $m\leq2$, $n\leq2$) and combining the DC Kerr nonlinearity takes the form
\begin{align}
    H_{\text{LINC}} &\approx \left(\omega_{\text{eff}}+2g_{22}\abs{\phi_{ac}}^2\right) \hat{c}^{\dagger} \hat{c} + \frac{1}{2}\alpha[\phi_{DC}] \hat{c}^{\dagger} \hat{c}^{\dagger} \hat{c} \hat{c} \nonumber\\&+ \left(2g_{21}\phi_{ac}(t) + 2g_{22}\phi_{ac}^2(t)\right) \left(\hat{c}^2 + \hat{c}^{\dagger2} \right),
\end{align}
where
\begin{align}
    \omega_{\text{eff}}&= \omega_0+ \alpha[\phi_{\text{DC}}], \\
    \omega_0 &= \sqrt{8E_c(E_L+2E_J\cos(\phi_{\text{DC}}))},\\
    \alpha[\phi_{\text{DC}}] &= -E_c \frac{2E_J\cos(\phi_{\text{DC}})}{E_L+2E_J\cos(\phi_{\text{DC}})}.
\end{align}

The flux-driven interaction produces single-mode squeezing, analogous to the three-wave-mixing process in charge-driven SPAs, with an effective coupling strength $g_{\text{3WM}}=2g_{21}$, where $g_{21}$ is generically given by
\begin{align}
    g_{21}
    &= \frac{E_L}{2}\frac{\partial \theta_{zpf}^2}{\partial \phi_d}\Big|_{\phi_{DC}} -2E_J\left[
        -\sin\phi_{DC}\left(-\tfrac{1}{2}\theta_{zpf}^2\right)
        \right. \nonumber\\ &\qquad \left.+ \cos\phi_{DC}\left(-\theta_{zpf}\,\frac{\partial \theta_{zpf}}{\partial \phi_d}\Big|_{\phi_{DC}}\right)
    \right] \nonumber\\
    &= \frac{E_L}{2}\theta_{zpf}^{-2}\frac{2E_CE_J\sin\phi_{DC}}{(E_L+2E_J\cos\phi_{DC})^2} \nonumber\\&-E_J \sin\phi_{DC}\,\theta_{zpf}^2 \,
       \frac{E_L+E_J\cos\phi_{\text{DC}}}{E_L+2E_J\cos\phi_{\text{DC}}},
\end{align}
based on Eq.~(\ref{gmn}).

Meanwhile, the flux drive induces a driven Zeeman shift $\Delta\omega=2g_{22}\abs{\phi_{ac}}^2$, where
\begin{align}
    g_{22}
    &= \frac{E_L}{4}\frac{\partial^2 \theta_{\text{zpf}}^2}{\partial^2 \phi_d}\Big|_{\phi_{\text{DC}}} -2E_J\left(-\frac{1}{2}\right)\cos\phi_{\text{DC}}\left(-\tfrac{1}{2}\theta_{zpf}^2\right) \nonumber \\& +E_J\cos\phi_{\text{DC}}\left[\theta_{zpf}\frac{\partial^2 \theta_{zpf}}{\partial \phi_d^2}\Big|_{\phi_{\text{DC}}} + \left(\frac{\partial \theta_{zpf}}{\partial \phi_d}\Big|_{\phi_{\text{DC}}}\right)^2\right]\nonumber\\ & -2E_J\left(-\frac{1}{2}\right)\sin\phi_{\text{DC}}\left(-2\theta_{zpf}\frac{\partial \theta_{zpf}}{\partial \phi_d}\Big|_{\phi_{\text{DC}}}\right).
\end{align}
By its definition, the Zeeman shift is observed in the lab frame under the static basis of a diagonalized, undriven Hamiltonian. It can be equivalently derived as a time-averaged AC Stark effect in the interaction frame via second-order perturbation theory; but its observable manifestation—a shift in the resonance frequency—is always referenced to the undriven static basis.

For LINCs which operate at $\phi_{\text{DC}}=\frac{\pi}{2}$, the parametric coefficients simplify as
\begin{align}
    g_{21}[\phi_{\text{DC}}=\pi/2]&=-\frac{E_J}{4E_L}\omega_L,\\
    g_{22}[\phi_{\text{DC}}=\pi/2]&=-\frac{E_J^2}{8E_L^2}\omega_L.
\end{align}

We now derive a comprehensive model for LPAs by including higher-order nonlinear effects under flux driving. Incorporating nonlinear terms up to $m\leq4$, $n\leq2$, which produce the AC-flux-driven Kerr nonlinearity and Stark shift, yields the Hamiltonian
\begin{align}
    H_{\text{LINC}} &= 4E_c\hat{n_c}^2\nonumber\\&+ \left(g_{20}+g_{21}\phi_{\text{ac}}(t)+g_{22}\phi_{\text{ac}}(t)^2\right)\left(\hat{c}+\hat{c}^{\dagger}\right)^2\nonumber\\&+\left(g_{40}+g_{41}\phi_{\text{ac}}(t)+g_{42}\phi_{\text{ac}}^2(t)\right)\left(\hat{c}+\hat{c}^{\dagger}\right)^4 \nonumber\\
    &=\left(\omega_{\text{eff}}+2g_{22}\abs{\phi_{\text{ac}}}^2\right) \hat{c}^{\dagger} \hat{c}+ \frac{1}{2}\alpha[\phi_{\text{DC}}] \hat{c}^{\dagger} \hat{c}^{\dagger} \hat{c} \hat{c} \nonumber\\&+ \left(2g_{21}\phi_{\text{ac}}(t) + 2g_{22}\phi_{\text{ac}}^2(t)\right) \left(\hat{c}^2 + \hat{c}^{\dagger2} \right)\nonumber\\&+(g_{41}\phi_{\text{ac}}(t)+g_{42}\phi_{\text{ac}}^2(t))\left(\hat{c}+\hat{c}^{\dagger}\right)^4\nonumber\\&\approx \left(\omega_{\text{eff}}+6g_{42}|\phi_{\text{ac}}|^2+2g_{22}\abs{\phi_{\text{ac}}}^2\right) \hat{c}^{\dagger} \hat{c}\nonumber\\ &+ \frac{1}{2}\left(\alpha[\phi_{\text{DC}}] +  6g_{42}|\phi_{\text{ac}}|^2\right) \hat{c}^{\dagger} \hat{c}^{\dagger} \hat{c} \hat{c} \nonumber\\&+ \left(2g_{21}\phi_{\text{ac}}(t) + 2g_{22}\phi_{ac}^2(t)\right) \left(\hat{c}^2 + \hat{c}^{\dagger 2} \right) \nonumber\\&
    +g_{41}\phi_{\text{ac}}(t)\Big(6\hat{c}^2+6\hat{c}^{\dagger2}\nonumber\\
    &\qquad+4\hat{c}^{\dagger}\hat{c}\hat{c}^{2}+4\hat{c}^{\dagger2}\hat{c}^{\dagger}\hat{c}\Big)\nonumber\\&+g_{42}\phi_{ac}^2(t)(c^4+c^{\dagger4}).\label{H_general}
\end{align}
The Hamiltonian provides a unified framework for describing SQUID- and LINC-based parametric amplifiers. Here the AC flux induces a photon-number-dependent dressed squeezing effect with the coefficient
\begin{align}
    g_{41}
    &= -2E_J \frac{1}{4!}\left[\cos\phi_{\text{DC}}\frac{\partial\tilde{\tilde{\theta}}_{\text{zpf}}}{\partial\phi_d}+
        \left(-\sin{\phi_{\text{DC}}}\right)\tilde{\tilde{\theta}}_{\text{zpf}}\right],
\end{align}
as well as an AC-flux-dependent stark shift depending on $g_{22}$ and $g_{42}$, where
\begin{align}
    g_{42}
    &= -2E_J \frac{1}{4!}\left[
        \frac{1}{2}\cos\phi_{\text{DC}}\frac{\partial^2\tilde{\tilde{\theta}}_{\text{zpf}}}{\partial\phi_d^2}-\sin{\phi_{\text{DC}}}\frac{\partial\tilde{\tilde{\theta}}_{\text{zpf}}}{\partial\phi_d}
        \right. \nonumber\\ &\qquad \left.-\frac{1}{2} \cos\phi_{\text{DC}}\tilde{\tilde{\theta}}_{\text{zpf}}
    \right],
\end{align}
with the definition $\tilde{\tilde{\theta}}_{\text{zpf}}\equiv\theta_{\text{zpf}}^4$.

For a LINC operating at $\phi_{\text{DC}}=\frac{\pi}{2}$, the Hamiltonian takes the form
\begin{align}
    H_{\text{LINC}}[\phi_{\text{DC}}=\frac{\pi}{2}]&=\left(\omega_{L}+2g_{22}\abs{\phi_{\text{ac}}}^2\right) \hat{c}^{\dagger} \hat{c}\nonumber\\&+ \left(2g_{21}\phi_{ac}(t)+2g_{22}\phi_{\text{ac}}^2(t)\right) \left(\hat{c}^2 + \hat{c}^{\dagger2} \right)\nonumber\\&+(g_{41}\phi_{\text{ac}}(t)+g_{42}\phi_{\text{ac}}^2(t))\left(\hat{c}+\hat{c}^{\dagger}\right)^4.
\end{align}
Applying an AC flux near $2\omega_L$ allows transformation of the Hamiltonian into the rotating frame under the RWA
\begin{align}
    H_{\text{LINC}} &= \hbar\delta_{\text{eff}} \hat{c}^{\dagger} \hat{c}+\frac{i \hbar}{2}\left( \epsilon_1 \hat{c}^{\dagger 2} -\epsilon_1^{*} \hat{c}^2\right)\nonumber\\&+\frac{i \hbar}{2}\left( \epsilon_2 \hat{c}^{\dagger 2} \hat{c}^{\dagger}\hat{c} -\epsilon_2^{*} \hat{c}^{\dagger}\hat{c} \hat{c}^2\right)\nonumber\\&+\frac{1}{2}\hbar K_{\text{eff}}\hat{c}^{\dagger 2} \hat{c}^{2}+\frac{1}{24}\hbar K_{\text{eff}}(e^{2i\phi_0}c^4+e^{-2i\phi_0}c^{\dagger4}),
\end{align}
where
\begin{align}
    \delta_{\text{eff}}&=\delta+6g_{42}|\phi_{ac}|^2=\Delta+2g_{22}\abs{\phi_{ac}}^2+6g_{42}|\phi_{ac}|^2,\\
    \epsilon_1&=-i(2g_{21}+6g_{41})|\phi_{ac}|e^{-i\phi_0},\\ \epsilon_2&=-4ig_{41}|\phi_{ac}|e^{-i\phi_0},\\ K_{\text{eff}}&=6g_{42}|\phi_{ac}|^2.
\end{align}
Additionally, the nonlinear coefficients for LINCs simplify as
\begin{align}
g_{41}[\phi_{DC}=\frac{\pi}{2}] &= \frac{E_J E_c}{6E_L} = \frac{E_J}{12\sqrt{2}E_L}\sqrt{\frac{E_c}{E_L}}\omega_L.\\
    g_{42}[\phi_{DC}=\frac{\pi}{2}] &= \frac{E_J^2 E_c}{3E_L^2} = \frac{E_J^2}{6\sqrt{2}E_L^2}\sqrt{\frac{E_c}{E_L}}\omega_L.
\end{align}

\section{Analysis of the LPA dynamics}\label{AppendixB}
To analyze the dynamics of a linear LPA, we transform Eqs.~(\ref{EM1}) and (\ref{EM2}) to the frequency domain
\begin{align}
-i\omega \hat{c}[\omega] &= \big(-i\delta - \frac{\gamma}{2}\big)\hat{c}[\omega] + \epsilon \{\hat{c}[-\omega]\}^{\dagger} \nonumber\\& + \sqrt{\gamma_c} \hat{a}_{\text{in}}[\omega]+\sqrt{\gamma_0} \hat{b}_{\text{in}}[\omega],\label{B1}
\end{align}
\begin{align}
-i\omega \{\hat{c}[-\omega]\}^{\dagger} &= \big(i\delta - \frac{\gamma}{2}\big)\{\hat{c}[-\omega]\}^{\dagger} + \epsilon^* \hat{c}[\omega] \nonumber\\& + \sqrt{\gamma_c} \{\hat{a}_{\text{in}}[-\omega]\}^{\dagger} +\sqrt{\gamma_0} \{\hat{b}_{\text{in}}[-\omega]\}^{\dagger},\label{B2}
\end{align}
where $\omega$ is the detuning between the signal frequency and $\frac{\omega_d}{2}$. By defining the Green's function
\begin{equation}
    G[\omega]= \begin{bmatrix}
    -i(\omega-\delta)+\frac{\gamma}{2} & -\epsilon\\
-\epsilon^* & -i(\omega+\delta)+\frac{\gamma}{2}
\end{bmatrix}^{-1},
\end{equation}
we can rewrite Eqs.~(\ref{B1})(\ref{B2}) as
\begin{align}
\begin{pmatrix}
\hat{c}[\omega]\\
\{\hat{c}[-\omega]\}^{\dagger}
\end{pmatrix}
&=  \sqrt{\gamma_c}\bm{G}[\omega]
\begin{pmatrix}
\hat{a}_{\text{in}}[\omega]\\ \{\hat{a}_{\text{in}}[-\omega]\}^{\dagger}\
\end{pmatrix} \nonumber\\
&+\sqrt{\gamma_0}\bm{G}[\omega]
\begin{pmatrix}
\hat{b}_{\text{in}}[\omega]\\\{\hat{b}_{\text{in}}[-\omega]\}^{\dagger}
\end{pmatrix},
\end{align}
Applying the input-output relation
\begin{equation}
\begin{pmatrix}
\hat{a}_{\text{out}}[\omega]\\
\{\hat{a}_{\text{out}}[-\omega]\}^{\dagger}
\end{pmatrix}
= \begin{pmatrix}
\hat{a}_{\text{in}}[\omega]\\
\{\hat{a}_{\text{in}}[-\omega]\}^{\dagger}
\end{pmatrix}
- \sqrt{\gamma_c}
\begin{pmatrix}
\hat{c}[\omega]\\
\{\hat{c}[-\omega]\}^{\dagger}
\end{pmatrix},
\end{equation}
we obtain
\begin{align}
\begin{pmatrix}
\hat{a}_{\text{out}}[\omega]\\
\{\hat{a}_{\text{out}}[-\omega]\}^{\dagger}
\end{pmatrix}
= &\bm{S}[\omega] \begin{pmatrix}
\hat{a}_{\text{in}}[\omega]\\
\{\hat{a}_{\text{in}}[-\omega]\}^{\dagger}
\end{pmatrix}
\nonumber\\&+ \bm{L}[\omega]
\begin{pmatrix}
\hat{b}_{\text{in}}[\omega]\\
\{\hat{b}_{\text{in}}[-\omega]\}^{\dagger}
\end{pmatrix},
\end{align}
with the scattering matrix of the signal
\begin{align}
\bm{S}[\omega] &= \bm{I} - \gamma_c \bm{G}[\omega],
\end{align}
and the scattering matrix for the intrinsic noise
\begin{equation}
\bm{L}[\omega] = -\sqrt{\gamma_0 \gamma_c} \bm{G}[\omega].
\end{equation}

Extending the above analysis to the comprehensive LINC model, we include the quartic nonlinear interactions of the near-resonance field components. In the semiclassical limit, the intracavity mean field can be decomposed into its frequency components as $\langle \hat{c} \rangle=\sum_{s,i,h}\alpha_{s,i,h}[\omega]e^{-i\omega_{s,i,h} t} $. Substituting this expansion into Eqs.~(\ref{EM3}) and (\ref{EM4}) yields the harmonic balanced equations
\begin{align}
    &(\omega-\delta_{\text{eff}}+i\gamma/2)\alpha_s =u_s+(2g_{\text{eff},1}+K_{\text{eff}}\alpha_h^2)\alpha_i^{*}\nonumber\\&+2g_{\text{eff},2}(2|\alpha_s|^2+|\alpha_i|^2+2|\alpha_h|^2)\alpha_i^{*}\nonumber\\&+\frac{1}{2}K_{\text{eff}}e^{-2i\phi_0}(\alpha_s^*\alpha_i^{*}+\alpha_h^{*2})\alpha_i^{*}\nonumber\\&+2g_{\text{eff,2}}[\alpha_h^{*2}+(\alpha_h^2+\alpha_s\alpha_i)e^{2i\phi_0}]\alpha_s\nonumber\\&+K_{\text{eff}}(|\alpha_s|^2+2|\alpha_i|^2+2|\alpha_h|^2)\alpha_s,
    \label{signal} \\[6pt]
    &(-\omega-\delta_{\text{eff}}+i\gamma/2)\alpha_i =u_i+(2g_{\text{eff},1}+K_{\text{eff}}\alpha_h^2)\alpha_s^{*}\nonumber\\&+2g_{\text{eff},2}(|\alpha_s|^2+2|\alpha_i|^2\nonumber+2|\alpha_h|^2)\alpha_s^{*}\nonumber\\&+\frac{1}{2}K_{\text{eff}}e^{-2i\phi_0}(\alpha_i^*\alpha_s^{*}+\alpha_h^{*2})\alpha_s^{*}\nonumber\\&+2g_{\text{eff,2}}[\alpha_h^{*2}+(\alpha_h^2+\alpha_s\alpha_i)e^{2i\phi_0}]\alpha_i\nonumber\\&+K_{\text{eff}}(2|\alpha_s|^2+|\alpha_i|^2+2|\alpha_h|^2)\alpha_i,
    \label{idler}\\[6pt]
    &(-\delta_{\text{eff}}+i\gamma/2)\alpha_h =u_h+(2g_{\text{eff},1}+2K_{\text{eff}}\alpha_s\alpha_i)\alpha_h^{*}\nonumber\\&+2g_{\text{eff},2}(2|\alpha_s|^2+2|\alpha_i|^2\nonumber+|\alpha_h|^2)\alpha_h^{*}\nonumber\\&+K_{\text{eff}}e^{-2i\phi_0}(\frac{1}{6}\alpha_h^{*2}+\alpha_{s}^{*}\alpha_{i}^{*})\alpha_h^{*}\nonumber\\&+4g_{\text{eff,2}}[\alpha_s^{*}\alpha_{i}^{*}+(\alpha_s\alpha_i+\frac{1}{6}\alpha_h^2)e^{2i\phi_0}]\alpha_h\nonumber\\&+K_{\text{eff}}(2|\alpha_s|^2+2|\alpha_i|^2+|\alpha_h|^2)\alpha_h,
\end{align}
where
\begin{align}
    g_{\text{eff},1}&=\frac{i}{2}\epsilon_1=(g_{21}+3g_{41})|\phi_{ac}|e^{-i\phi_0},\\
    g_{\text{eff},2}&=\frac{3i}{4}\epsilon_2=3g_{41}|\phi_{ac}|e^{-i\phi_0}.
\end{align}

With no input at $\omega_d/2$, $\alpha_h=0$ is a steady-state solution. In the zero-signal limit, linearizing about this solution gives
\begin{equation}
    \lambda_{\pm}=-\frac{\gamma}{2}\pm\sqrt{4|g_{\mathrm{eff},1}|^2-\delta_{\mathrm{eff}}^2}.
\end{equation}
The zero-field solution is stable when
\begin{equation}
    4|g_{\mathrm{eff},1}|^2<\delta_{\mathrm{eff}}^2+\frac{\gamma^2}{4}.
    \label{eq:linear_stability}
\end{equation}
We choose the small-signal operating point to satisfy this condition and consider amplification on the $\alpha_h=0$ branch below the parametric oscillation (PO) threshold.

Considering the regime under the PO threshold where $\alpha_h=0$, one can rewrite Eqs.~(\ref{signal}-\ref{idler}) via a susceptibility matrix
\begin{align}
\begin{pmatrix}
\alpha_s\\
\alpha_i^{*}
\end{pmatrix}
&= \frac{1}{D}
\begin{pmatrix}
-\omega-\Delta_s-i\gamma/2 & 2g_{\text{eff},s}\\ 2g_{\text{eff},i}^* & \omega-\Delta_i+i\gamma/2
\end{pmatrix}
\begin{pmatrix}
u_s\\u_i^{*}
\end{pmatrix},
\end{align}
where
\begin{align}
    g_{\text{eff,i}}&=g_{\text{eff},1}+g_{eff,2}(|\alpha_s|^2+2|\alpha_i|^2\nonumber)+\frac{1}{4}K_{\text{eff}}e^{-2i\phi_0}\alpha_s^*\alpha_i^{*}\\&=\left[g_{21}+3g_{41}(1+|\alpha_s|^2+2|\alpha_i|^2)\right]|\phi_{\text{ac}}|e^{-i\phi_0}\nonumber\\&+\frac{3}{2}g_{42}|\phi_{\text{ac}}|^2e^{-2i\phi_0}\alpha_s^*\alpha_i^{*},\label{geffi}\\
    g_{\text{eff,s}}&=g_{\text{eff},1}+g_{\text{eff},2}(2|\alpha_s|^2+|\alpha_i|^2\nonumber)+\frac{1}{4}K_{\text{eff}}e^{-2i\phi_0}\alpha_s^*\alpha_i^{*}\\&=\left[g_{21}+3g_{41}(1+2|\alpha_s|^2+|\alpha_i|^2)\right]|\phi_{\text{ac}}|e^{-i\phi_0}\nonumber\\&+\frac{3}{2}g_{42}|\phi_{\text{ac}}|^2e^{-2i\phi_0}\alpha_s^*\alpha_i^{*},\label{geffs}
\end{align}
\begin{align}
    \Delta_s&=\delta_{\text{eff}}+K_{\text{eff}}(2|\alpha_s|^2+|\alpha_i|^2)+2g_{\text{eff,2}}^*e^{-2i\phi_0}\alpha^*_s\alpha^*_i\\&=\Delta+2g_{22}|\phi_{ac}|^2+6g_{42}|\phi_{ac}|^2\nonumber\\&+6g_{42}|\phi_{\text{ac}}|^2(2|\alpha_s|^2+|\alpha_i|^2)+6g_{41}|\phi_{ac}|e^{-i\phi_0}\alpha^*_s\alpha^*_i,\label{Delta_s}
\end{align}
\begin{align}
    \Delta_i&=\delta_{\text{eff}}+K_{\text{eff}}(|\alpha_s|^2+2|\alpha_i|^2)+2g_{\text{eff,2}}e^{2i\phi_0}\alpha_s\alpha_i\\&=\Delta+2g_{22}|\phi_{\text{ac}}|^2+6g_{42}|\phi_{\text{ac}}|^2\nonumber\\&+6g_{42}|\phi_{\text{ac}}|^2(|\alpha_s|^2+2|\alpha_i|^2)+6g_{41}|\phi_{\text{ac}}|e^{i\phi_0}\alpha_s\alpha_i,\label{Delta_i}
\end{align}
\begin{align}
    D&=-\omega^2+\Delta_s \Delta_i + \omega(\Delta_i-\Delta_s)+\gamma^2/4-4g_{\text{eff,i}}^*g_{\text{eff,s}}\nonumber\\&-i\gamma(\omega+\frac{\Delta_s-\Delta_i}{2}).
\end{align}
The last terms in Eqs.~(\ref{geffi}) and (\ref{geffs}) can be neglected because $|\phi_{\text{ac}}|^2\ll|\phi_{\text{ac}}|$. Meanwhile, the last terms in Eqs.~(\ref{Delta_s}) and (\ref{Delta_i}) generate the nonlinear corrections to the complex detuning. In the near-resonant, weak-signal limit, the last term in Eq.~(\ref{Delta_s}) is approximately
\begin{equation}
    \frac{24ig_{41}g_{21}}{\gamma}|\phi_{\mathrm{ac}}|^2|\alpha_s|^2\equiv i\gamma_{\Delta},
\end{equation}
which only contributes to an intensity-dependent linewidth narrowing.

Using the input-output relation $u_{\text{out}}=i\gamma(a-a^{\dagger})-u_{\text{in}}$~\cite{frattini2018optimizing}, the total output field can be expressed as
\begin{align}
    u_{out}&=i\gamma(\alpha_s e^{-i\omega_s t}+\alpha_i e^{-i\omega_i t} - \alpha_s^* e^{i\omega_s t} - \alpha_i^* e^{i\omega_i t}) \nonumber\\&- u_s e^{-i\omega_s t} - u_i e^{-i\omega_i t}.
\end{align}
The output field contains an amplified signal component proportional to $e^{-i\omega t}$
\begin{align}
    u_{out}^{s} & =i\gamma\alpha_s - u_s \nonumber\\&=\frac{i\gamma}{D}\left(-\omega-\Delta_s-\frac{i\gamma}{2}\right) u_s-u_s.
\end{align}
We can thus derive the full expression of the power-dependent gain
\begin{align}
    G & =\abs{\frac{i\gamma}{D}\left(-\omega-\Delta_s-\frac{i\gamma}{2}\right) -1}^2.
\end{align}

Under the small-signal approximation, the gain simplifies to
\begin{align}
    G_0 =1+\frac{4\gamma^2\abs{g_{\text{eff}}}^2}{\left[-\omega^2+\delta_{\text{eff}}^2+\frac{\gamma^2}{4}-4\abs{g_{\text{eff}}}^2\right]^2+\gamma^2\omega^2},
\end{align}
where
\begin{align}
    g_{\text{eff}}=\left(g_{21}+3g_{41}\right)|\phi_{\text{ac}}|e^{-i\phi_0},
\end{align}

Beyond the small-signal regime, the gain depends on the numbers of signal and idler photons within the bosonic mode of the LPA. For large gain, the photon numbers are approximated as $n_s\approx n_i \approx \frac{GP_{\text{in}}}{\hbar\omega_L \gamma}$. The relation between $G$ and $P_{\text{in}}$ can be found by numerically solving the nonlinear equations:

\begin{equation}
\tilde{g}_{\text{eff}}(n_s) = \left(g_{21} + 3g_{41} + 9g_{41} \frac{\tilde{G} P_{\text{in}}}{\hbar \omega_L \gamma}\right)\abs{\phi_{\text{ac}}}e^{-i\phi_0},
\label{eq:geff}
\end{equation}

\begin{align}
|\tilde{D}|^2 &= \left[ -\omega^2 + \tilde{\Delta}^2 + \frac{(\gamma+2\gamma_{\Delta})^2}{4} - 4\abs{\tilde{g}_{\text{eff}}}^2 \right]^2 \nonumber\\&+ (\gamma+2\gamma_{\Delta})^2 \omega^2,
\label{eq:D1}
\end{align}

\begin{equation}
\tilde{G} = 1 + \frac{
\begin{gathered}
4\gamma(\gamma+2\gamma_{\Delta})|\tilde{g}_{\text{eff}}|^2\\
-2\gamma\gamma_{\Delta}
\left[(\omega+\tilde{\Delta})^2+\frac{(\gamma+2\gamma_{\Delta})^2}{4}\right]
\end{gathered}
}{|\tilde{D}|^2},
\label{eq:G}
\end{equation}

\begin{align}
\tilde{\Delta} &= \Delta + \left(2g_{22} + 6g_{42}\right)|\phi_{\text{ac}}|^2+ 18g_{42} \frac{\tilde{G}P_{\text{in}}}{\hbar\omega_L \gamma}
|\phi_{\text{ac}}|^2.
\label{eq:Deltas}
\end{align}

\section{Intermodulation of LPA}\label{AppendixC}

We calculate the third-order intermodulation distortion (IMD)
by considering the nonlinear processes induced by the driven Kerr
nonlinearity and dressed squeezing, including the coupling
between the generated IM sidebands and their conjugate
components through linear squeezing.

The IMD is typically characterized by the JPA's response to two near-resonance input fields. Assuming they are at the frequencies
$\omega_{s_1}$ and $\omega_{s_2}$, which, in the rotating frame, become $\delta_{s_1}$ and $\delta_{s_2}$, with $\delta_{s_j}=\omega_{s_j}-\omega_d/2$ ($j=1,2$). The output field contains
the amplified signals at $\delta_{s_1,s_2}$, their idlers at $-\delta_{s_1,s_2}$, and the generated
third-order IM sidebands:
\begin{align}
&2\delta_{s_1}-\delta_{s_2},\quad
2\delta_{s_2}-\delta_{s_1},\nonumber\\
&-2\delta_{s_1}+\delta_{s_2},\quad
-2\delta_{s_2}+\delta_{s_1}.
\end{align}

To derive the IMD property, we solve Eq.~(\ref{EM3}) by expanding the intracavity field as
\begin{align}
c(t)={}&
\alpha_{s_1}e^{-i\delta_{s_1}t}
+\alpha_{s_2}e^{-i\delta_{s_2}t}
\nonumber\\
&+\beta_{s_1}e^{i\delta_{s_1}t}
+\beta_{s_2}e^{i\delta_{s_2}t}
\nonumber\\
&+\alpha_{A1}^{\text{IM}}
e^{-i(2\delta_{s_1}-\delta_{s_2})t}
\nonumber\\
&+\alpha_{B1}^{\text{IM}}
e^{i(2\delta_{s_1}-\delta_{s_2})t}
+\cdots,
\label{IIP_field}
\end{align}
Here, $\alpha_{s_j}$ and $\beta_{s_j}$ denote the amplitudes of the signal
and the idler, respectively. They can be derived from the linearized EOM by retaining the linear squeezing $\varepsilon_1\hat c^\dagger$ and neglecting
the photon-number-dependent nonlinear terms. Meanwhile, the amplitudes $\alpha_{A1}^{\text{IM}}$ and
$\alpha_{B1}^{\text{IM}}$ describe the generated
third-order sidebands at
$2\delta_{s_1}-\delta_{s_2}$ and
$-2\delta_{s_1}+\delta_{s_2}$,
respectively.

From the linearized EOM, the signal and idler
amplitudes satisfy
\begin{align}
0={}&
\left[
-i(\delta_{\text{eff}}-\delta_{s_j})
-\frac{\gamma}{2}
\right]\alpha_{s_j}
+\varepsilon_1\beta_{s_j}^*
+\sqrt{\gamma_c}\,a_{{\rm in},s_j},
\\
0={}&
\left[
-i(\delta_{\text{eff}}+\delta_{s_j})
-\frac{\gamma}{2}
\right]\beta_{s_j}
+\varepsilon_1\alpha_{s_j}^*.
\end{align}
Consequently,
\begin{equation}
\beta_{s_j}^*=r_j\alpha_{s_j},
\qquad
r_j=
\frac{\varepsilon_1^*}
{\gamma/2-i(\delta_{\text{eff}}+\delta_{s_j})}.
\label{IIP_idler_relation}
\end{equation}

For two closely spaced input signals, we approximate
$r_1\simeq r_2\simeq r$, such that
\begin{equation}
\beta_{s_j}^*=r\alpha_{s_j}.
\label{IIP_r}
\end{equation}

We now consider the EOMs of the IM sideband at
$2\delta_{s_1}-\delta_{s_2}$ and its conjugate
component at $-(2\delta_{s_1}-\delta_{s_2})$
\begin{align}
\frac{d\alpha_{A1}^{\text{IM}}}{dt}
={}&
\left[
-i(\delta_{\text{eff}}
-2\delta_{s_1}+\delta_{s_2})
-\frac{\gamma}{2}
\right]\alpha_{A1}^{\text{IM}}
\nonumber\\
&+\varepsilon_1\alpha_{B1}^{\text{IM}*}
+C_A\alpha_{s_1}^2\alpha_{s_2}^*,
\label{A1_EOM}
\\
\frac{d\alpha_{B1}^{\text{IM}}}{dt}
={}&
\left[
-i(\delta_{\text{eff}}
+2\delta_{s_1}-\delta_{s_2})
-\frac{\gamma}{2}
\right]\alpha_{B1}^{\text{IM}}
\nonumber\\
&+\varepsilon_1\alpha_{A1}^{\text{IM}*}
+C_B^*\alpha_{s_1}^{*2}\alpha_{s_2}.
\label{B1_EOM}
\end{align}
Here the nonlinear coefficients $C_A$ and $C_B$ include all the third-order IM processes generated by Kerr
nonlinearity and dressed squeezing:
\begin{align}
C_A={}&
-iK_{\text{eff}}(1+2|r|^2)-\frac{i}{2}K_{\text{eff}}e^{-2i\phi_0}r^2
\nonumber\\
&+\frac{3}{2}\varepsilon_2 r(2+|r|^2)-\frac{3}{2}\varepsilon_2^*r^*.
\label{IIP_CA}
\end{align}

\begin{align}
C_B={}&
iK_{\text{eff}}r(2+|r|^2)+\frac{i}{2}K_{\text{eff}}e^{2i\phi_0}r^*
\nonumber\\
&+\frac{3}{2}\varepsilon_2^*(1+2|r|^2)-\frac{3}{2}\varepsilon_2 r^2.
\label{IIP_CB}
\end{align}
Taking the complex conjugate of
Eq.~(\ref{B1_EOM}) and imposing the steady-state
condition, the coupled equations can be written as
\begin{equation}
M_{\text{IM}}
\begin{bmatrix}
\alpha_{A1}^{\text{IM}}\\
\alpha_{B1}^{\text{IM}*}
\end{bmatrix}
=
\begin{bmatrix}
C_A\\
C_B
\end{bmatrix}
\alpha_{s_1}^2\alpha_{s_2}^*,
\label{IIP_matrix}
\end{equation}
where
\begingroup
\setlength{\arraycolsep}{5pt}
\begin{equation}
M_{\text{IM}}=
\begin{bmatrix}
\begin{gathered}
i(\delta_{\text{eff}}-2\delta_{s_1}+\delta_{s_2})\\
+\gamma/2
\end{gathered}
&-\varepsilon_1\\
-\varepsilon_1^*
&\begin{gathered}
-i(\delta_{\text{eff}}+2\delta_{s_1}-\delta_{s_2})\\
+\gamma/2
\end{gathered}
\end{bmatrix}.
\end{equation}
\endgroup
The inverse matrix is
\begingroup
\setlength{\arraycolsep}{5pt}
\begin{align}
M^{-1}_{\text{IM}}&=
\frac{1}{D_{\text{IM}}}\times\nonumber\\&
\begin{bmatrix}
\begin{gathered}
-i(\delta_{\text{eff}}+2\delta_{s_1}-\delta_{s_2})\\
+\gamma/2
\end{gathered}
&\varepsilon_1\\
\varepsilon_1^*
&\begin{gathered}
i(\delta_{\text{eff}}-2\delta_{s_1}+\delta_{s_2})\\
+\gamma/2
\end{gathered}
\end{bmatrix},
\end{align}
\endgroup
where
\begin{align}
D_{\text{IM}}={}&
\delta_{\text{eff}}^2
-(2\delta_{s_1}-\delta_{s_2})^2
+\frac{\gamma^2}{4}
\nonumber\\
&-i\gamma(2\delta_{s_1}-\delta_{s_2})
-|\varepsilon_1|^2.
\label{IIP_D}
\end{align}
Thus, the generated IM sideband amplitude can be expressed as
\begin{align}
\alpha_{A1}^{\text{IM}}
={}&
\frac{1}{D_{\text{IM}}}
\Bigg\{
\left[
\frac{\gamma}{2}
-i(\delta_{\text{eff}}
+2\delta_{s_1}-\delta_{s_2})
\right]C_A\nonumber\\
&\qquad+\varepsilon_1C_B
\Bigg\}
\alpha_{s_1}^2\alpha_{s_2}^*.
\label{IIP_amplitude}
\end{align}

Using the input-output relation
\begin{equation}
a_{\rm out}=a_{\rm in}-\sqrt{\gamma_c}\,c,
\end{equation}
and noting that no external field is injected at
the IM frequency, we obtain the output field amplitude
\begin{equation}
a_{\rm out}^{s,\text{IM}}
=-\sqrt{\gamma_c}\,
\alpha_{A1}^{\text{IM}}.
\end{equation}
The corresponding output power is
\begin{equation}
P_{\text{out}}^{s,\text{IM}}
=\hbar\omega_{A1}\gamma_c
\left|\alpha_{A1}^{\text{IM}}\right|^2,
\end{equation}
Substituting Eq.~(\ref{IIP_amplitude}), we obtain
\begin{align}
P_{\text{out}}^{s,\text{IM}}
={}&
\hbar\omega_{A1}\gamma_c
\Bigg|
\frac{1}{D_{\text{IM}}}
\Bigg\{
\left[
\frac{\gamma}{2}
-i(\delta_{\text{eff}}
+2\delta_{s_1}-\delta_{s_2})
\right]C_A
\nonumber\\
&\qquad+\varepsilon_1C_B
\Bigg\}
\Bigg|^2
n_{s_1}^2n_{s_2},
\label{IIP_output_general}
\end{align}
where $n_{s_j}=|\alpha_{s_j}|^2$.
For a strongly overcoupled, single-port reflection amplifier
($\gamma_c\simeq\gamma$), in the on-resonance, narrow-detuning limit used below, the intracavity signal
photon number is related to the input power by
\begin{equation}
n_{s_j}\hbar\omega_{s_j}\gamma
\simeq
(\sqrt{G}+1)^2P_{\text{in}}^{s_j},
\label{IIP_gain_relation}
\end{equation}
where $G=G_0$ in the weak-signal regime.
For two closely spaced signals with equal input
powers,
\begin{equation}
P_{\text{in}}^{s_1}
=P_{\text{in}}^{s_2}
=P_{\text{in}}^s,
\qquad
n_{s_1}\simeq n_{s_2},
\end{equation}
and $\omega_{s_1}\simeq\omega_{s_2}
\simeq\omega_{A1}\simeq\omega_L$,
the IM output power becomes
\begin{align}
P_{\text{A1,out}}^{\text{IM}}
={}&
\hbar\omega_L\gamma
\Bigg|
\frac{1}{D_{\text{IM}}}
\Bigg\{
\left[
\frac{\gamma}{2}
-i(\delta_{\text{eff}}
+2\delta_{s_1}-\delta_{s_2})
\right]C_A
\nonumber\\
&\qquad+\varepsilon_1C_B
\Bigg\}
\Bigg|^2
\times
\left[
\frac{(\sqrt{G}+1)^2}
{\hbar\omega_L\gamma}
P_{\text{in}}^s
\right]^3.
\label{IIP_output_power}
\end{align}

If we set $\delta_{\text{eff}}=0$ by adjusting
the pump frequency and choose the two signals
near resonance, such that
$2\delta_{s_1}-\delta_{s_2}\approx0$,
the output IMD power simplifies to
\begin{align}
P_{\text{A1,out}}^{\text{IM}}
=
\hbar\omega_L\gamma
\left|
\frac{
\frac{\gamma}{2}C_A+\varepsilon_1C_B
}{D_{\text{IM}}}
\right|^2
\left[
\frac{(\sqrt{G}+1)^2}
{\hbar\omega_L\gamma}
P_{\text{in}}^s
\right]^3,
\label{IIP_output_resonance}
\end{align}
where
\begin{equation}
D_{\text{IM}}=\frac{\gamma^2}{4}-|\varepsilon_1|^2.
\end{equation}
Imposing $P_{\text{A1,out}}^{\text{IM}}=P_{\text{S1,out}}$ with $G=G_0$ yields $\text{IIP}_3$
\begin{align}
    \text{IIP}_3 = \left|\frac{\frac{\gamma^2}{4}-|\varepsilon_1|^2}{\frac{\gamma}{2}C_A+\epsilon_1C_B}\right|\frac{\sqrt{G_0}}{(\sqrt{G_0}+1)^3}\hbar\omega_L \gamma.
\end{align}

\section{SQUID-based parametric amplifier}\label{AppendixD}
A superconducting quantum interference device (SQUID) consists of a superconducting loop interrupted by Josephson junctions with intrinsic capacitance, enabling flux-tunable inductance and nonlinear interactions. Unlike a LINC, a SQUID lacks a shunting linear inductance requiring a DC flux bias away from $\pi/2$ to maintain a finite resonance frequency. The SQUID analysis therefore follows the general framework developed in \hyperref[AppendixA]{Appendix A}, with $E_L=0$ and an appropriately chosen DC flux bias $\phi_{\mathrm{DC}}$. Thus, the quadratic parametric coefficients become
\begin{align}
    g_{21}^S[E_L^S=0] &= -\frac{1}{2}E_J^S \sin\phi_{\mathrm{DC}}\theta_{zpf}^2\nonumber\\&=-\frac{1}{2}E_J^S \sin\phi_{\mathrm{DC}}\left(\frac{E_C^S}{E_J^S\cos\phi_{\mathrm{DC}}}\right)^{1/2}\nonumber\\&=-\frac{1}{8}\tan\phi_{\mathrm{DC}}\omega_S,
\end{align}
\begin{align}
g_{22}^S[E_L^S=0]
&=
-\frac{\sqrt{E_C^S E_J^S}}{8}
\frac{1+\cos^2\phi_{\mathrm{DC}}}
{\cos^{3/2}\phi_{\mathrm{DC}}}\nonumber\\&=
-\frac{1}{32}
\left(1+\sec^2\phi_{\mathrm{DC}}\right)\omega_S,
\end{align}
with the resonance frequency $\omega_S=4\sqrt{E_C^S E_J^S \cos\phi_{\text{DC}}}$.
Importantly, in contrast to LINCs, the SQUID retains a finite static Kerr nonlinearity with the coefficient
\begin{equation}
    \alpha^S[\phi_{\mathrm{DC}}]=12g_{40}^S=-E_C^S,
\end{equation}
In the instantaneous oscillator basis, the quartic potential
coefficient is independent of flux for $E_L^S=0$, giving
\begin{align}
    g_{41}^S[E_L^S=0] &= 0,
    \\ g_{42}^S[E_L^S=0] &= 0.
\end{align}

Building on the above analysis, we derive the SQUID Hamiltonian from the general expression in Eq.~(\ref{H_general}). Setting $E_L^S=0$ and substituting the nonlinear coefficients derived above, the Hamiltonian simplifies to
\begin{align}
    H_{\text{SQUID}}[E_L^S=0]&=\left(\omega^S_{\text{eff}}+2g_{22}^S\abs{\phi_{ac}}^2\right) \hat{c}^{\dagger} \hat{c}\nonumber\\&+ \frac{1}{2}\alpha^S[\phi_{DC}] \hat{c}^{\dagger} \hat{c}^{\dagger} \hat{c} \hat{c} \nonumber\\&+ \left(2g_{21}^S\phi_{ac}(t) + 2g_{22}^S\phi_{ac}^2(t)\right) \left(\hat{c}^2 + \hat{c}^{\dagger 2} \right),
\end{align}
under the definition $\omega^S_{\text{eff}}=\omega_S+\alpha^S[\phi_{DC}]$. The Hamiltonian can be further written in the form of a standard parametric amplifier
\begin{align}
H_{\mathrm{SQUID}}
=&\hbar\delta^S_{\mathrm{eff}}\hat c^\dagger\hat c
+\frac{i\hbar}{2}
\left(\epsilon_1^S\hat c^{\dagger 2}
-\epsilon_1^{S*}\hat c^2\right)+\frac{\hbar}{2}K_{\mathrm{eff}}^S
\hat c^{\dagger 2}\hat c^2,
\end{align}
where
\begin{align}
\delta^S_{\mathrm{eff}}
&=\omega_S - \omega_d/2+2g_{22}^S|\phi_{\mathrm{ac}}|^2-E_C^S,\\
\epsilon_1^S
&=-2ig_{21}^S|\phi_{\mathrm{ac}}|e^{-i\phi_0},\\
K_{\mathrm{eff}}^S
&=-E_C^S.
\end{align}

Following the LPA analysis, the field components in a SQUID-based JPA satisfy the harmonic-balanced equations
\begin{align}
    &(\omega-\delta^S_{\text{eff}}+i\gamma/2)\alpha_s =u_s+(2g^S_{\text{eff}}+K^S_{\text{eff}}\alpha_h^2)\alpha_i^{*}\nonumber\\&+K^S_{\text{eff}}(|\alpha_s|^2+2|\alpha_i|^2+2|\alpha_h|^2)\alpha_s, \\
    &(-\omega-\delta^S_{\text{eff}}+i\gamma/2)\alpha_i =u_i+(2g^S_{\text{eff}}+K^S_{\text{eff}}\alpha_h^2)\alpha_s^{*}\nonumber\\&+K^S_{\text{eff}}(2|\alpha_s|^2+|\alpha_i|^2+2|\alpha_h|^2)\alpha_i,
\end{align}
where $g^S_{\text{eff}}=\frac{i}{2}\epsilon^S_1$.
Solving the above equations yields the power-dependent gain
\begin{align}
    G^S & =\abs{\frac{i\gamma}{D^S}\left(-\omega-\Delta^S_s-\frac{i\gamma}{2}\right) -1}^2 =1+\frac{4\gamma^2|g^S_{\text{eff}}|^2}{\abs{D^S}^2},
\end{align}
where
\begin{align}
    \Delta^S_s&=\delta^S_{\text{eff}}+K^S_{\text{eff}}(2|\alpha_s|^2+|\alpha_i|^2),
\end{align}
\begin{align}
    \Delta^S_i&=\delta^S_{\text{eff}}+K^S_{\text{eff}}(|\alpha_s|^2+2|\alpha_i|^2),
\end{align}
\begin{align}
    \abs{D^S}^2&=\gamma^2\left(\omega+\frac{\Delta^S_s-\Delta^S_i}{2}\right)^2\nonumber\\&+\left[\Delta^S_s \Delta^S_i+\omega(\Delta^S_i-\Delta^S_s)-\omega^2+\frac{\gamma^2}{4}-4|g^S_{eff}|^2\right]^2.
\end{align}
Furthermore, the IMD property can be analyzed following the approach presented in \hyperref[AppendixC]{Appendix C}. Here for the SQUID, $\text{IIP}_3$ takes the form
\begin{align}
    \text{IIP}_3 = \left|\frac{\frac{\gamma^2}{4}-|\varepsilon^S_1|^2}{\frac{\gamma}{2}C^S_A+\epsilon^S_1C^S_B}\right|\frac{\sqrt{G^S_0}}{(\sqrt{G^S_0}+1)^3}\hbar\omega_S \gamma,
\end{align}
where
\begin{align}
C^S_A=&
-iK^S_{\text{eff}}(1+2|r^S|^2),\\
C^S_B=&
iK^S_{\text{eff}}r^S(2+|r^S|^2),\\
r^S=&
    \frac{\varepsilon^{S*}_1}
{\gamma/2}.
\end{align}
It is worth noting that the conventional analytical treatment of third-order IMD in parametric amplifiers approximates the nonlinear drive using the direct Kerr mixing of the two signal tones. This approximation neglects additional third-order mixing processes involving the parametrically generated idlers and the conjugate intermodulation sideband, which can make nontrivial contribution in the high-gain regime.

To ensure a fair comparison between the SQUID and LINC
parametric amplifiers, we consider identical resonance
frequencies, charging energies, and three-wave-mixing
coefficients. Imposing $\omega_L=\omega_S=\omega_0$ gives
\begin{equation}
    E_J^S =
    \frac{E_L}{2\cos\phi_{\mathrm{DC}}^S}.
\end{equation}
We further require identical three-wave-mixing
coefficients, $|g_{21}|=|g_{21}^S|$, yielding
\begin{equation}
    E_J =
    \frac{E_L}{2}\tan\phi_{\mathrm{DC}}^S.
\end{equation}
Choosing $\phi_{\mathrm{DC}}^S=\pi/3$,
the circuit parameters satisfy
\begin{align}
    E_C^S &= E_C,\\
    E_L &= \frac{\omega_0^2}{8E_C},\\
    E_J^S &= E_L,\\
    E_J &= \frac{\sqrt{3}}{2}E_L.
\end{align}
In addition to the above requirement of the circuit parameters, the AC flux amplitudes are independently adjusted to
achieve the same small-signal gain of $20\,\mathrm{dB}$,
while the resonance linewidths are kept identical.

\section{SNAIL parametric amplifier}\label{AppendixE}
A superconducting nonlinear asymmetric inductive element (SNAIL) consists of a superconducting loop containing three identical large Josephson junctions in one branch and a smaller one in the other. The SNAIL is biased by an external flux, allowing control of its resonance frequency and nonlinear potential. In addition to the strongly coupled signal port, a weakly coupled port delivers the current pump required to produce wave-mixing processes. The Hamiltonian of the lowest mode of a SNAIL parametric amplifier (SPA) can be written in the form \cite{sivak2019kerr}
\begin{align}
    H_{\text{SPA}}/\hbar = \omega_a(\phi)\hat{a}^\dagger \hat{a}+\sum_{n=3}^{\infty} g_{n}[\phi]\left(\hat{a}+\hat{a}^{\dagger}\right)^n.
\end{align}
The asymmetric SNAIL potential supports a cubic interaction for current-pumped three-wave mixing. In the effective model below, the static-Kerr cancellation condition is $g_4-5g_3^2/\omega_a=0$, as follows from Eq.~(\ref{app_Kerr}) at $n_p=0$.

Applying a current pump at frequency $\omega_p=2(\omega_a+\Delta)$, we produce an intracavity displacement with mean amplitude $\alpha_p$, which triggers the three-wave mixing process for parametric amplification. The near-resonance behavior of the SPA can be described by an effective Hamiltonian
\begin{align}
    H_{\text{eff}}/\hbar &= - \Delta_b \hat{b}^\dagger \hat{b} +  ( g \hat{b}^{\dagger2} + g^* \hat{b}^2)+\frac{K}{2}\hat{b}^{\dagger 2}\hat{b}^2,
\end{align}
where
\begin{align}
    \Delta_b (n_p) &= \Delta-(\frac{32}{3}g_4-28\frac{g_3^2}{\omega_a})n_{p},\\
    g(\alpha_p)&=2g_3\alpha_p,\\
    K(n_p)&=12(g_4-5\frac{g_3^2}{\omega_a})-\frac{256}{3}\frac{g_4^2 n_p}{\omega_a}.\label{app_Kerr}
\end{align}

Following the harmonic balance approach, we can obtain the steady-state equations for signals and idlers up to the second order
\begin{align}
&\Biggl[
\omega_s - \omega_a + i \frac{\kappa}{2}\nonumber\\
&
- g_4 \left(
\frac{32}{3} |\alpha_p|^2 + 12 |\alpha_s|^2 + 24 |\alpha_i|^2 + 24 |\alpha_h|^2
\right)\nonumber\\
&+ 4\frac{g_3^2}{\omega_a} \left[
7 |\alpha_p|^2 + 15 |\alpha_s|^2 + 36 |\alpha_i|^2
\right] \nonumber\\&+ \frac{256}{3}\frac{g_4^2 |\alpha_p|^2}{\omega_a}\left[|\alpha_s|^2+2|\alpha_i|^2+2|\alpha_h|^2\right]
\Biggr]\alpha_s\nonumber\\
&= u_s + 4 g_3 \alpha_p \alpha_i^{*},\\[6pt]\nonumber
\end{align}
\begin{align}
    &\Biggr[
\omega_i - \omega_a + i \frac{\kappa}{2}\nonumber\\
&
- g_4 \Bigg(
\frac{32}{3} |\alpha_p|^2 + 12 |\alpha_i|^2 + 24 |\alpha_s|^2 + 24 |\alpha_h|^2
\Biggl)\nonumber\\
&+ 4\frac{g_3^2}{\omega_a} \left[
7 |\alpha_p|^2 + 36 |\alpha_s|^2 + 15 |\alpha_i|^2
\right] \nonumber\\&+ \frac{256}{3}\frac{g_4^2 |\alpha_p|^2}{\omega_a}\left[2|\alpha_s|^2+|\alpha_i|^2+2|\alpha_h|^2\right]
\Biggr]\alpha_i
\nonumber\\&= u_i + 4 g_3 \alpha_p \alpha_s^{*}.
\end{align}
Transforming the above equations into the susceptibility matrix form yields
\begin{align}
\begin{pmatrix}
\alpha_s\\
\alpha_i^{*}
\end{pmatrix}
&= \frac{1}{D}
\begin{pmatrix}
-\omega-\Delta_s-i\kappa/2 & 2g_{\text{eff}}\\ 2g_{\text{eff}}^* & \omega-\Delta_i+i\kappa/2
\end{pmatrix}
\begin{pmatrix}
u_s\\u_i^{*}
\end{pmatrix},
\end{align}
where
\begin{align}
    g_{\text{eff}}=2g_3\alpha_p,
\end{align}
\begin{align}
    \Delta_{s,i} = &\Biggl(
\omega_a-\frac{\omega_p}{2} \nonumber\\&+g_4 \left[
\frac{32}{3} |\alpha_p|^2 + 12 |\alpha_{s,i}|^2 + 24 |\alpha_{i,s}|^2 + 24 |\alpha_h|^2
\right]\nonumber\\
&- 4\frac{g_3^2}{\omega_a} \left[
7 |\alpha_p|^2 + 15 |\alpha_{s,i}|^2 + 36 |\alpha_{i,s}|^2
\right] \nonumber\\&- \frac{256}{3}\frac{g_4^2 |\alpha_p|^2}{\omega_a}\left[|\alpha_{s,i}|^2+2|\alpha_{i,s}|^2+2|\alpha_h|^2\right]
\Biggr),
\end{align}
and
\begin{align}
    D&=-\omega^2+\Delta_s \Delta_i + \omega(\Delta_i-\Delta_s)+\kappa^2/4-4\abs{g_{\mathrm{eff}}}^2\nonumber\\&-i\kappa(\omega+\frac{\Delta_s-\Delta_i}{2}).
\end{align}

\bibliographystyle{apsrev4-2}
\bibliography{references}\end{document}